\documentclass[a4paper,twoside,12pt]{article}

\usepackage{graphicx}
\usepackage{psfig}
\usepackage{amsfonts}
\usepackage{amssymb}

\newcommand{\ZZ}{\mathbb{Z}}

\title{Simulations of Alice Electrodynamics on a Lattice }
\author{J.Striet$^1$ and F.A.Bais$^2$\\[2mm]
Institute for Theoretical Physics \\ University of Amsterdam \\
Valckenierstraat 65 \\ 1018XE Amsterdam \\ The Netherlands\\}

\begin{document}
\maketitle
\date
\footnotetext[1]{jelpers@science.uva.nl}
\footnotetext[2]{bais@science.uva.nl}


\begin{abstract}
\noindent In this paper we present results of numerical simulations
and some (analytical) approximations of a compact $U(1)\ltimes\ZZ_2$
lattice gauge theory, including an extra bare mass term for Alice
fluxes. The subtle interplay between Alice fluxes and (Cheshire)
magnetic charges is analysed. We determine the phase diagram and some
characteristics of the model in three and four dimensions. The results
of the numerical simulations in various regimes, compare well with
some analytic approximations.
\end{abstract}

\section{Introduction}
In this paper we investigate a lattice version of Alice
electrodynamics (LAED). Alice electrodynamics (AED) is a gauge theory
with gauge group $U(1)\ltimes\ZZ_2\sim O(2)$, in other words a
minimally non-abelian extension of ordinary electrodynamics
\cite{schwarz}. The nontrivial $\ZZ_2$ transformation reverses the
direction of the electric and magnetic fields and the sign of the
charges. This means that in Alice electrodynamics charge conjugation
symmetry is gauged. However, as this non-abelian extension is
discrete, it only affects electrodynamics through certain global
(topological) features, such as the appearance of Alice fluxes (or
vortices) and Cheshire charges \cite{alford}. The topological
structure of $U(1)\ltimes\ZZ_2$ differs from that of $U(1)$ in a few
subtle points. Firstly, AED allows topologically stable vortices since
$\Pi_0(U(1)\ltimes\ZZ_2) = \ZZ_2$, these will be referred to as Alice
fluxes. Note that in this theory this flux is co\"existing with the
unbroken $U(1)$ of electromagnetism and therefore it is not an
``ordinary'' magnetic flux. Secondly, just as the compact $U(1)$ gauge
theory, AED also contains magnetic monopoles, because
$\Pi_1(U(1)\ltimes\ZZ_2) = \ZZ$. We note however, that due to the fact
that the $\ZZ_2$ and the $U(1)$ part of the gauge group do not
commute, magnetic charges of opposite sign belong to the same
topological class. The aim of this paper is to get an understanding of
the phase diagram in a simple lattice version of Alice electrodynamics
but which does contain both monopoles and Alice fluxes. We do so by
simulations and some analytic approximations.

The paper is organised as follows. In section \ref{model} we specify
the lattice model in detail. In section \ref{phasediagram} we give the
numerical results we obtained for the phase diagrams of the model in
dimensions three and four. In section \ref{approximations} we present
some analytic approximations related to the phase diagram and other
measurable quantities. In the final section we summarise the results
and draw conclusions.

\section{Lattice Alice Electrodynamics}
\label{model}
In this section we introduce a specific LAED model. First we explain
the different terms that appear in the action and then we discuss how
magnetic monopoles and instantons are realized and can be
measured. Finally we say a few things about the computational
implementation of the model.

\subsection{The action}
Alice phases can be generated by spontaneously breaking $SU(2)$ to
$U(1)\ltimes\ZZ_2$. In this case it is clear that Alice loops,
monopoles and Cheshire charges may arise as regular classical finite
energy solutions. In the study presented here we restrict ourselves to
compact $U(1)\ltimes\ZZ_2$ gauge theory with an extra bare mass term
for the Alice fluxes. Our lattice formulation of the theory allows for
the formation of Alice fluxes and magnetic monopoles\footnote{It also
allows for the formation of Cheshire charges, but their non-locality
makes them hard to detect, see section \ref{moninlaed}}. The action we
will use is given by:
\begin{equation}
I = \frac{1}{g^2} \sum_{plaquettes}\{-\Re(Tr(U_1 U_2 U_3^\dagger
U_4^\dagger))+ m_f P_f\}~.
\label{action}
\end{equation}
The first part represents the normal Wilson \cite{wilson} plaquette
action for the gauge theory. The second term is the extra bare mass
term for the $\ZZ_2$ fluxes in the model. $P_f$ is a functional of the
$\ZZ_2$ degrees of freedom which, when evaluated on a plaquette,
equals one if the plaquette is pierced by a $\ZZ_2$ flux, and equals
zero if not. The parameter $m_f$ is the extra bare mass (in three
dimensions) or tension (in four dimensions) for the Alice
flux.

In principle one can also add an extra bare monopole mass term to the
action.  We have refrained from doing so because it is computationally
much more involved and because we can realize all four phases
in the model without this term (see table~\ref{phases}). To define
suitable link variables for LAED we use the fact that compact
$U(1)\ltimes\ZZ_2$ can be conveniently embedded in $SU(2)$, leading
to:
\begin{equation}
U_\nu(x) = e^{i A_\nu(x) \tau_3} \tau_1^{a_\nu(x)}~,
\label{link}
\end{equation}
with $a_\nu(x)\in\{0,1\}$ and $A_\nu(x)\in\langle-\pi,\pi]$. Thus
$a_\nu$ represents the $\ZZ_2$ gauge variable and $A_\nu$ the compact
$U(1)$ gauge variable of the theory. We say that, if $a_\nu(x)=1$ a
$\ZZ_2$-sheet in 3D, or a $\ZZ_2$-volume in 4D, crosses the link,
implying that the $\ZZ_2$-sheets live on the dual lattice. These
$\ZZ_2$-sheets can, of course, be moved around by local $\ZZ_2$ gauge
transformations. A gauge transformation of the links is given by:
\begin{equation}
U_\nu(x)\to\Omega(x)U_\nu(x)\Omega(x+\hat{\nu})^\dag~,
\label{gaugetrans}
\end{equation}
with $\Omega(x)=e^{i \Sigma_\nu(x) \tau_3} \tau_1^{\sigma_\nu(x)}$,
where $\sigma_\nu(x)\in\{0,1\}$ and
$\Sigma_\nu(x)\in\langle-\pi,\pi]$.\\ The boundaries of the
$\ZZ_2$-sheets, however, cannot be moved around by local $\ZZ_2$ gauge
transformations, this in analogy with the endpoints of Dirac strings
(being magnetic monopoles) in the compact $U(1)$ gauge theory. This is
exactly what one should expect, since the boundaries of the
$\ZZ_2$-sheets are closed Alice flux\footnote{To avoid confusion we
note that in the $\ZZ_n$ literature one typically calls these objects
vortices instead of fluxes.} loops, which are physical objects
carrying energy. Bearing this in mind it is easy to locate the Alice
fluxes, namely by just counting the number of $\ZZ_2$-sheets crossing
the links of a plaquette. If an even number of $\ZZ_2$-sheets crosses
the links of the plaquette, then no Alice flux pierces the plaquette,
but if an odd number does, then that means that an Alice flux does
pierce the plaquette. This observation allows one to define the $P_f$
operator, which applied to a plaquette measures the presence of an
Alice flux,
\begin{equation}
P_f=\frac{1}{2}(1-(-1)^{\sum_{i=1}^{4} a_i})~,
\label{Pflux}
\end{equation}
where the four $a_i$ summed over belong to links $U_1$, $U_2$, $U_3$
and $U_4$, bounding a single plaquette. Equations (\ref{action}),
(\ref{link}) and (\ref{Pflux}) define our LAED model.

\subsection{The problem of locating monopoles (or instantons)}
\label{moninlaed}
In this model of LAED in four dimensions, there are magnetic
monopoles, in three dimensions these appear as instantons. There are a
few intricacies in detecting them compared to the usual compact $U(1)$
lattice gauge theory. In this section we will explain under what
circumstances and how we can detect a monopole/instanton in LAED.
As our model of LAED has a lot of similarities with compact $U(1)$
lattice gauge theory, we try to  use these similarities in
determining the monopole content of a configuration.

Let us first consider the case that there are no Alice fluxes
present. Clearly, this corresponds to the limit of an infinitely large
mass, $m_f$, for the flux. In this case there may still be closed
$\ZZ_2$ surfaces, but these surfaces are not physical and can be moved
around by making suitable local $\ZZ_2$ transformations.  Suppose we
want to determine the monopole content of a specific cube in such a
configuration. We would like to see if a Dirac string ends in the
cube, just as one does for compact $U(1)$ lattice gauge theory. We 
distinguish two cases, the first where no $\ZZ_2$-sheet crosses the
cube of interest and the second where one or more $\ZZ_2$-sheets do
cross the cube.

In the first case we determine the monopole content of the cube just
as in compact $U(1)$ lattice gauge theory. In the second case we
should find a new or more general definition due to the presence of
the $\ZZ_2$-sheets. Bearing in mind that a monopole is a
physical object which cannot be moved around by gauge transformations,
one may use local $\ZZ_2$ gauge transformations to gauge the
$\ZZ_2$-sheets away from the cube of interest. After this procedure we
can again determine the monopole content by the methods of compact $U(1)$
lattice gauge theory.

$\ZZ_2$-sheets can be gauged out of the cube of interest in several
different ways. One would expect this not to make any
difference to the outcome of the measurement of the monopole charge of
the cube of interest, but it does!  As we
mentioned in the introduction, monopoles of opposite sign belong to
the same topological class. For the measurement of the monopole charge
of a single cube this means that one cannot distinguish between
positive and negative charges. To see that this is the case, let us
consider a cube which is not intersected by a $\ZZ_2$-sheet. If one
performs a 'global' $\ZZ_2$ gauge transformation to all the links of
this cube, this has the same effect as pulling a $\ZZ_2$-sheet through
the cube; all the $U(1)$ degrees of freedom change sign, since $\tau_1
e^{iA\tau_3}\tau_1=e^{-iA\tau_3}$, see equation (\ref{gaugetrans}). Obviously this means that the outcome of the
measurement of the magnetic charge of the cube changes sign. Hence
only the absolute value of the magnetic charge is a locally gauge
invariant quantity, i.e. an observable.

Next we consider the situation where fluxes are present. Now we have
two different type of cubes, cubes which are pierced by a flux and
cubes which are not. The latter are obviously equivalent to the cubes
we just discussed. Thus at this point we may restrict our
considerations to cubes which are pierced by fluxes.  The statement
is, that for a cube which is pierced by a flux, the notion of a gauge
invariant magnetic charge breaks down completely. Let us explain why
this is the case.

\begin{figure}[!htb]
\begin{center}
\includegraphics[width=7.9cm,height=6cm,clip]{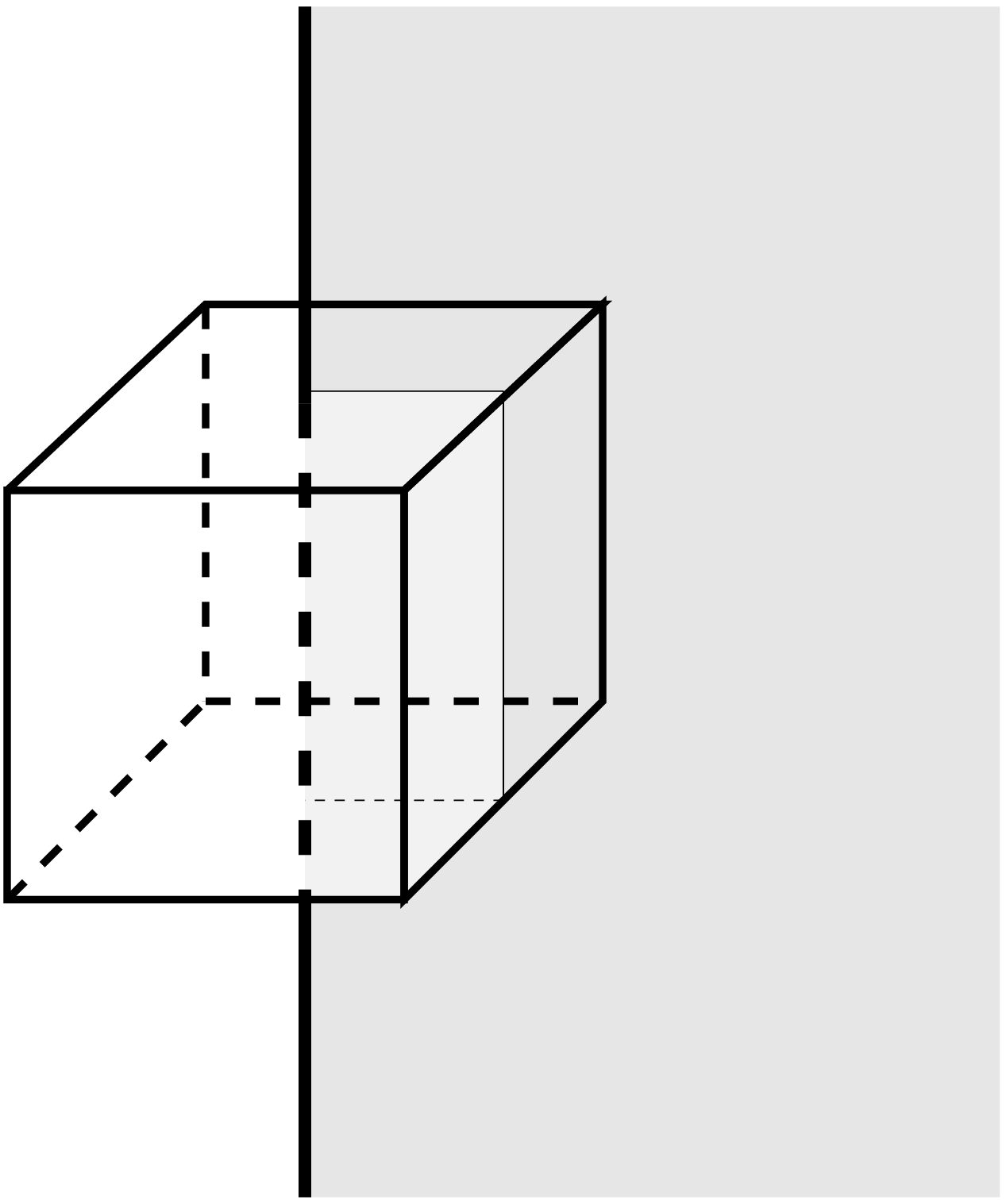}
\caption[somethingelse]{\footnotesize A cube that is pierced by a flux
which is the boundary of a $\ZZ_2$-sheet.}
\label{piercube.eps}
\end{center}
\end{figure}

\noindent If an Alice flux pierces through a cube, it is obviously not
possible to gauge the $\ZZ_2$-sheet out of the cube. In figure
\ref{piercube.eps} we depicted a cube pierced by a flux and the
$\ZZ_2$-sheet connected to the flux. If one tries to define Dirac
strings through the plaquettes bounding the cube of interest one gets
into all sorts of trouble. For the plaquettes where no flux pierces
through one can up to a sign determine the (real magnetic flux through
the) Dirac string. This sign problem seems to be a minor one, as it
appears to be for the monopole charge itself, but that is not true,
because there is a separate sign ambiguity for the Dirac strings
through each of the plaquettes and not just a single overall sign, as
was the case for a cube not pierced by a flux. This means that in such
a cube, even the absolute value of the net magnetic charge is not an
invariant quantity.

Yet another problem arises if one wants to define the Dirac string
through a plaquette which is pierced by a flux, because an odd number
of $\ZZ_2$-sheets cross the links bounding the plaquette. The problem
basically follows directly from Alice electrodynamics itself, where if
one sweeps a $\ZZ_2$-sheet through a $U(1)$ link field, this will
change sign.  So even the sign of the individual $U(1)$ link variables
is not defined uniquely on a plaquette which is intersected by an odd
number of sheets, {\em even if one looks only at that particular
plaquette}.  This obstruction to defining the magnetic flux through
such a plaquette, is just a manifestation of what is generally called
the obstruction to globally define a $U(1)$ charge in the presence of
an Alice flux in Alice electrodynamics.

However not all is lost. The previous discussion
only shows that it is impossible to determine the magnetic
charge of a cube, or more general of a volume, whose bounding surface is
pierced by a flux. There is however no problem in determining the
magnetic charge of a volume which contains a loop of flux not
crossing the boundary.
\begin{figure}[!htb]
\begin{center}
\includegraphics[width=7.9cm,clip]{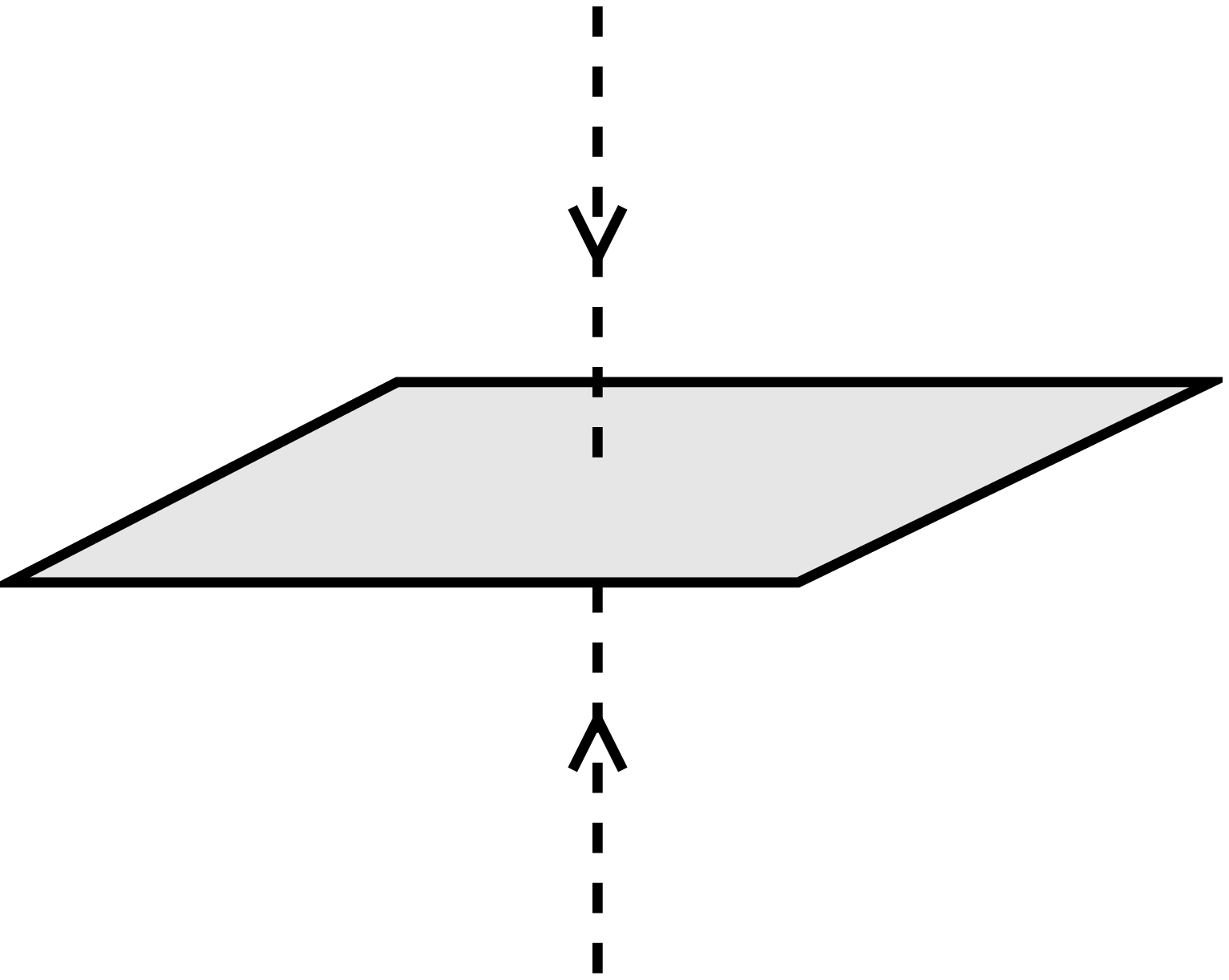}
\caption[somethingelse]{\footnotesize This figure shows an Alice 
loop with its $\ZZ_2$-sheet. The $\ZZ_2$-sheet is pierced by a Dirac
string, which changes sign/direction once it passes the
$\ZZ_2$-sheet. In this configuration the Alice loop carries a magnetic
Cheshire charge.}
\label{chescharge.eps}
\end{center}
\end{figure}

\noindent
To see this consider the configuration given in figure
\ref{chescharge.eps}.  A configuration is shown of an Alice loop and a
Dirac string piercing the $\ZZ_2$-sheet bounded by the Alice
loop. This figure demonstrates, that an Alice loop configuration is
capable of carrying a magnetic charge. We note that there is no Dirac
string coming from the flux itself (this is actually possible and even
necessary for the unit charged Alice loop). Remember that we are, for
plaquettes not pierced by a flux, able to determine the Dirac string
up to a sign. We also note that any attempt to measure the location of
the monopole will fail. It looks like that the cube where the the
Dirac string pierces the $\ZZ_2$-sheet does contain a magnetic charge,
but as the position of the sheet is gauge dependent this is just a
gauge illusion. Yet, drawing a closed 2-surface around the loop there
is a gauge invariant quantity of magnetic flux emanating from that
surface, i.e. there is magnetic charge inside.  This magnetic charge
is present, not as a localised or even localiseable quantity, but
rather as a global property carried by a closed Alice loop as a whole,
in which case one speaks of a magnetic ``Cheshire'' charge. And
indeed, although one can determine the magnetic charge carried by the
Alice loop as a whole, one can not assign this magnetic charge to any
of the cubes inside the volume containing the loop. These
nonlocaliseable charges may in the continuum even be energetically
favoured, as we showed in \cite{jelper3}, 't Hooft Polyakov monopoles
may decay in their Cheshire versions.

We conclude, that once we enter a phase where there are very many Alice
fluxes around, detecting and localising magnetic charge gets a hairy
business. The only useful thing one may still do, is to measure the
fraction of monopole carrying cubes of the number of cubes {\em not}
pierced by an Alice flux. In view of these observations, when in the
following we talk about the monopole density, we mean the average
absolute charge per unpierced cube and when we talk about flux density
we mean the fraction of plaquettes pierced by an Alice flux,
i.e. $\langle P_f\rangle$, unless stated otherwise.

\subsection{Implementation of the model}
\label{implem}
Although formula (\ref{link}) suggests that we should implement LAED
using (Pauli) matrices we have not done so.  Instead, we exploited the
fact that the structure of our $U(1)\ltimes\ZZ_2$ gauge theory is very
close to that of the compact $U(1)$.  The only effect of the $\ZZ_2$
degrees of freedom is the appearance of Alice fluxes and
$\ZZ_2$-sheets. If there are an odd number of $a$ variables equal to
one in a plaquette, then the plaquette is pierced by a $\ZZ_2$ flux
and the first term in the action is always zero irrespective of the
values of the $A$ fields. This can be understood as a consequence of
the fact that the $U(1)$ symmetry is globally frustrated in the
presence of an Alice flux. If, in the contrary, there are an even or
zero number of $a$ variables equal to one in a plaquette, the $a$
variables can be gauged away, changing only the sign of some of the
$A$ fields and the action is just the action of compact $U(1)$. In
view of these observations, we have for our simulations used the
following simple action, which is equivalent to the action of formula
(\ref{action}), but does not require any matrix calculations.
\begin{equation}
I=\frac{1}{g^2} \sum_{plaquettes}\{-(1-P_f)\cos{\tilde{F}} + m_f
P_f\}~,
\label{action2}
\end{equation}
where $\tilde{F}$ is the $F$ of $U(1)$ after the $\ZZ_2$ fields have
been gauge transformed out of the plaquette, which is always possible
if $P_f=0$.

We have investigated this model, using a combination of a Monte Carlo
method for the $A_\nu(x)$ variables, and a heat bath method for the
$a_\nu(x)$ variables. We examined the model on a periodic hyper cubic
lattice of size $10^d$, where d is the dimension. Although we will not
go into detail on the order of the phase transitions we mention that
it has been suggested \cite{lang}, that the order, oddly enough,
would depend on the imposed boundary conditions.

Our LAED model contains a pure compact $U(1)$ and a $\ZZ_2$ gauge
theory in different limits of the model. In the limit of
$m_f\to\infty$ the model is equal to pure compact $U(1)$ gauge
theory. In the limit of $g^2\to\infty$ while keeping $m_f/g^2$ finite
the model is equal to $\ZZ_2$ gauge theory.  Before we proceed we like
to mention a few things about the $\ZZ_2$ gauge theory to avoid
confusion later on. In $\ZZ_2$ gauge theory there is only one
parameter, in the $\ZZ_2$ limit of our model this parameter is
$m_f/g^2$. Normally, the $\ZZ_2$ gauge theory is only studied for
positive values of its parameter. However, in our situation we are
also interested in the region where $m_f/g^2$ becomes negative.  In
the pure $\ZZ_2$ gauge theory the region of positive and negative
values of the parameter form a mirror image of each other. Note that
this mirror map is different from the usual duality that is also
present in $\ZZ_n$ type gauge theories.  This mirror symmetry holds,
at least, for a hyper cubic lattice, where one may map the negative
coupling side on the positive side if one replaces ``fluxes'' by
``no-fluxes'' in every sense. So ``no-fluxes'' are the places where
``no flux'' pierces through a plaquette, i.e. they are the holes in
the flux condensate. The model can equally well be described by either
of the two objects.  This mirror symmetry follows from the fact that
for a hyper cubic lattice both objects, fluxes and no-fluxes, form
closed loops in three dimensions and closed surfaces in four
dimensions.  This shows that the regions of positive values and
negative values of $m_f/g^2$ are can be naively mapped onto each
other.  As we will show, in LAED the Alice mirror symmetry is broken
by the interactions with the $U(1)$ gauge fields for finite values of
$g^2$.

\section{The phase diagram in three and four dimensions}
\label{phasediagram}
In this section we present various numerical results for the LAED
model. Because we have two types of topological objects in the theory,
which may or may not condense, one may in principle expect four
phases. It is quite easy to anticipate where in the parameter space
the four phases could occur, as we have indicated in table
\ref{phases}.

\begin{figure}[!htb]
\mbox{\psfig{figure=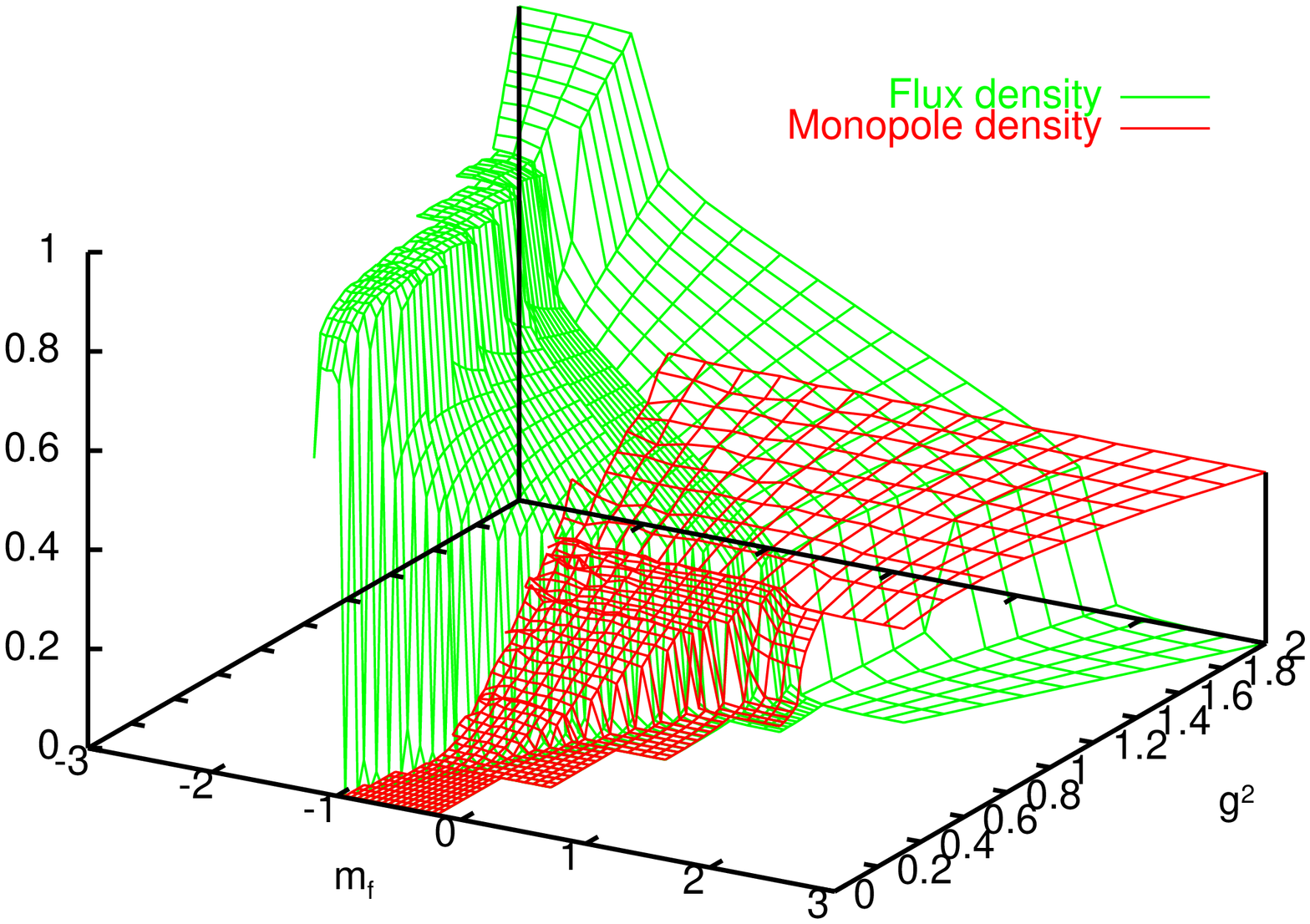,angle=270,width=7.9cm}}
\mbox{\psfig{figure=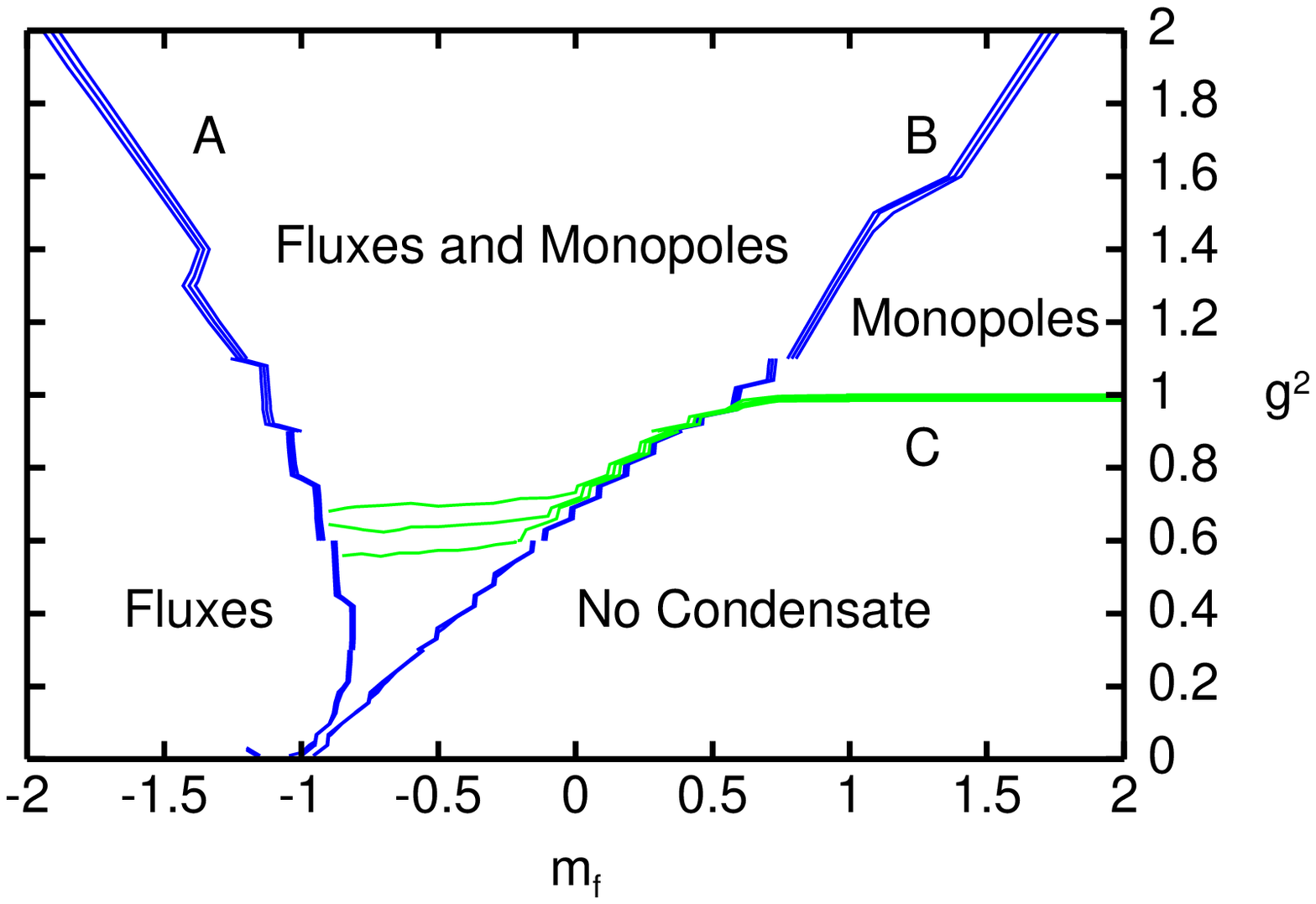,angle=270,width=7.9cm}}
\makebox[7.9cm][l]{\footnotesize{(a)}}
\makebox[7.9cm][l]{\footnotesize{(b)}}
\caption[somethingelse] 
{\footnotesize \\ (a): The 4-dimensional flux and the monopole
densities are plotted as a function of $m_f$ and $g^2$. The four
different phases of table \ref{phases} can be clearly distinguished.\\
(b): A plot of some specific monopole and the Alice flux density
contours in four dimensions. We identify the four phases of the
model. The lines denoted $B$ mark the transition involving the Alice
fluxes, where to the left of $B$ the fluxes are condensed. The lines
$A$ correspond to a second phase transition involving the fluxes. The
lines $C$ denote the monopole transition, notice the splitting of the
height lines once the Alice fluxes are condensed.}
\label{phase4d.ps}
\label{phase4dtop.ps}
\end{figure}

\begin{table}[!htb]
\centerline{
\begin{tabular}{|c||c|c|}\hline
& & \\ & $m_f$ small& $m_f$ large\\ & & \\
\hline
& & \\ $g^2$ small&Fluxes&No Condensate\\ & & \\ $g^2$ large&Fluxes
and Monopoles/&Monopoles/\\ & Instantons&Instantons \\
& & \\\hline
\end{tabular}
}
\caption[somethingelse]{The four phases of LAED.}
\label{phases}
\end{table} 
In figure \ref{phase4d.ps}(a) we have plotted the flux density and the
monopole density in four dimensions.  It is clear that various
interesting transitions do occur. Using a hysteresis type of analysis
we could determine the order of these transitions, and we found that
all but one, are of first order. Only the transition from the phase
with only Alice fluxes condensed, to the phase where both Alice fluxes
and monopoles are condensed, is different. In fact, it does appear not
to be a phase transition at all, but rather a crossover phenomenon,
see also section \ref{monopole} and the discussion in section
\ref{disofres}.

In figure \ref{phase4dtop.ps}(b) we have plotted some contours for the
Alice flux and monopole densities. The curves indicate where the
first order phase transitions take place, but also show the change of
the first order monopole transition if no fluxes are condensed, to the
crossover monopole transition if fluxes are condensed.

Note that in figures \ref{phase4d.ps}(a) and \ref{phase4dtop.ps}(b) we
have only plotted the monopole density up to the 'second' flux density
transition, line A in figure \ref{phase4dtop.ps}(b), where the flux
density jumps to about one and only very few cubes (if any) are left
where no flux pierces through, making the fluctuations for the
monopole measurement very large.

Though in our limited model we find all of the anticipated four
phases, each characterised by some condensate, we do not find all
possible transitions from one phase to another. There is apparently no
transition from the phase with condensed monopoles and no fluxes to
the phase with condensed fluxes and no monopoles.

In appropriate limits of the model we recover the results for the
lattice gauge theories of compact $U(1)$ and $\ZZ_2$ separately,
consistent with equation (\ref{action2}). The pure $U(1)$ gauge theory
arises in the limit of $m_f\to\infty$, where the Alice fluxes are
suppressed and the only feature reminiscent of the $\ZZ_2$ part of the
gauge theory are pure $\ZZ_2$ gauge transformations, which of course
do not affect any of the physics. In this limit we therefore expect
only the transition corresponding to monopole condensation. The pure
$\ZZ_2$ gauge theory arises in the limit of $g^2\to\infty$, while
keeping $m_f/g^2$ finite, which is usually only studied with $m_f/g^2
\geq 0$. We verified that the limiting behaviours of the results of
our simulations are in agreement with the known results of the $\ZZ_2$
and $U(1)$ gauge theories \cite{Balian,bhanot1,creutz1,degrand}, see
also
\cite{klaus} and references therein.
\begin{figure}[!htb]
\mbox{\psfig{figure=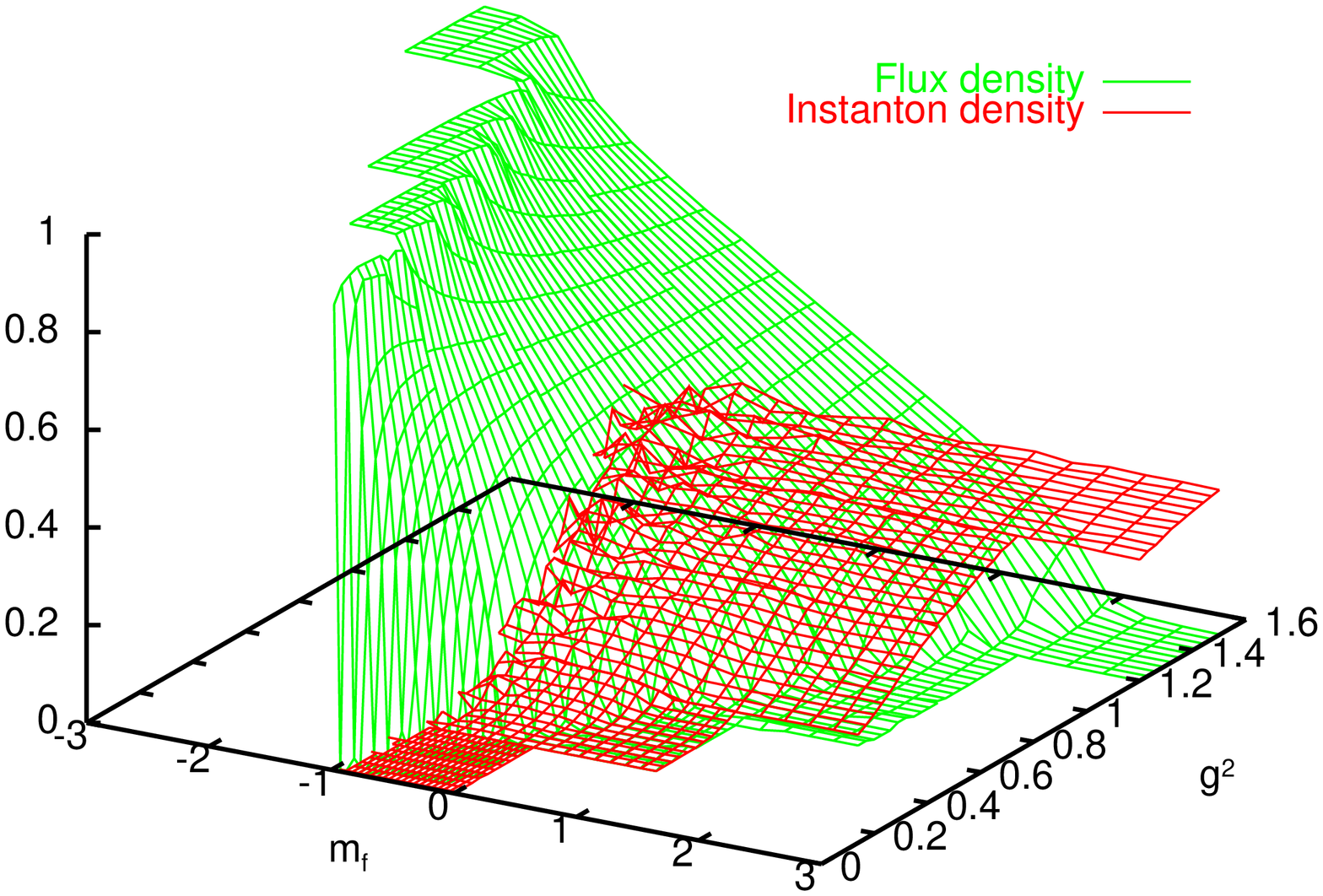,angle=270,width=7.9cm}}
\mbox{\psfig{figure=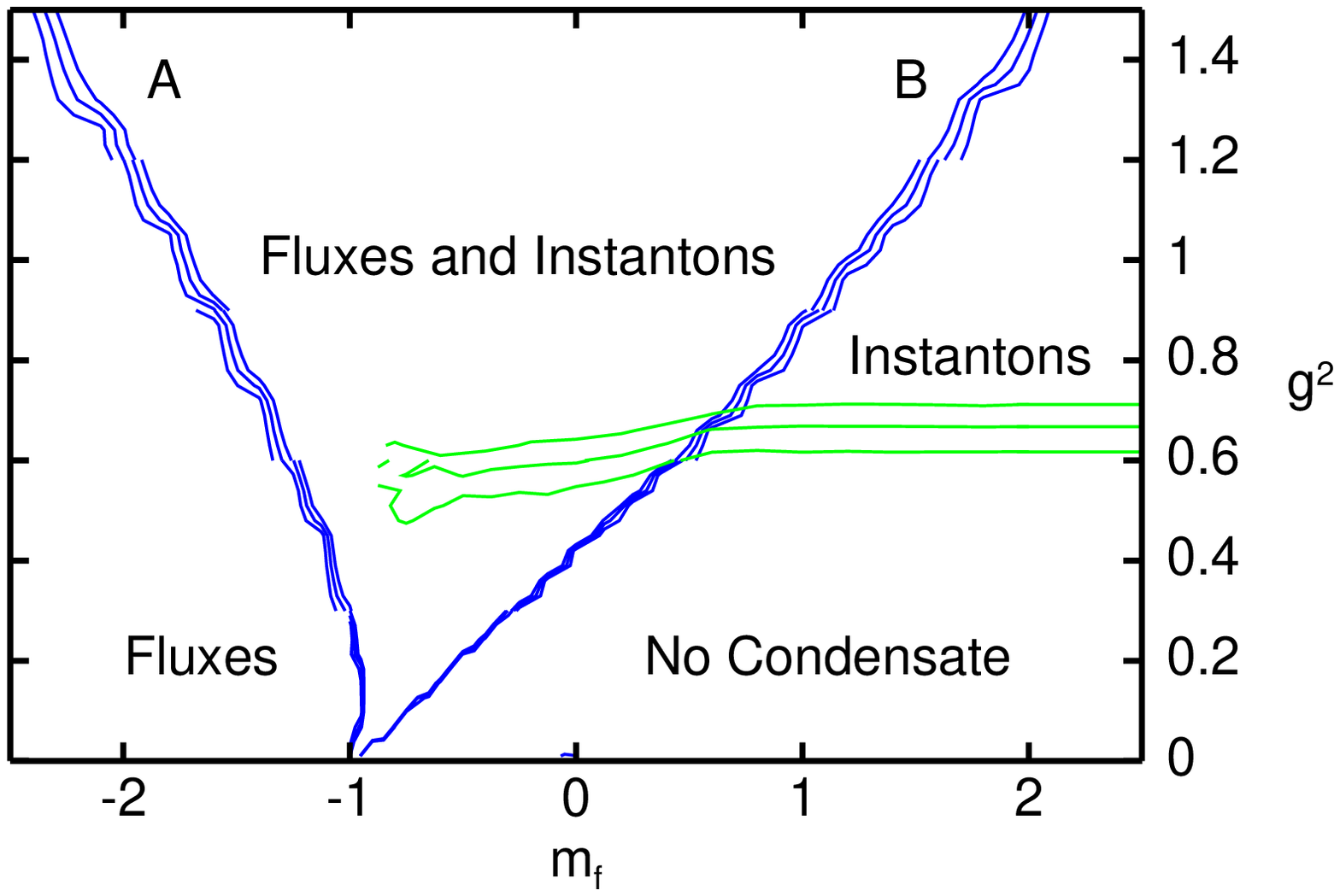,angle=270,width=7.9cm}}
\makebox[7.9cm][l]{\footnotesize{(a)}}
\makebox[7.9cm][l]{\footnotesize{(b)}}
\caption[somethingelse] 
{\footnotesize \\(a): The 3-dimensional flux and instanton densities
are plotted as a function of $m_f$ and $g^2$. The four different
phases of table \ref{phases} are clearly distinguishable.\\ (b): In
this figure we plotted some specific height lines of the instanton and
the Alice flux density in three dimensions. We identify the four
phases of the model. Line $B$ is the condensation line of the Alice
fluxes, to the left of it the fluxes are condensed. Line $A$ is a
second phase transition. In comparison with figure
\ref{phase4dtop.ps}(b) there is no line $C$. In three dimensions the
instanton condensation is always a crossover.}
\label{phase3d.ps}
\label{phase3dtop.ps}
\end{figure}

In figure \ref{phase3d.ps}(a) we plotted the results for the instanton
and Alice flux density in three dimensions.  Also in this case we
encounter all four phases of the theory, but the transitions are of
different order. The instanton condensation is always a crossover and
the flux condensation appears to be of second order, which it
certainly should be in the $\ZZ_2$ gauge theory limit
\cite{bhanot1}. We did not determine the order of the flux
condensation for small $g^2$.

In figure \ref{phase3d.ps}(a) the transition for small values of $g^2$
appears to become a first order phase transition, but this is mainly
due to the fact that we use $m_f$ and $g^2$ to parameterise the model,
whereas the, in some sense more natural, choice of $(m_f+1)/g^2$ and
$1/g^2$ could give a different picture, which is also true for the
four dimensional case. We will come back to this point in section
\ref{disofres}.

In figure \ref{phase3dtop.ps}(b) we, just as in figure
\ref{phase4dtop.ps}(b), plotted specific height lines of the
instanton and Alice flux density. These lines show the location of the
Alice flux phase transitions and divide the parameter space up in the
four different regions linked to the phases. Again in the $U(1)$ and
$\ZZ_2$ limit we recover the results of these pure gauge theories
separately.

In three dimensions the flux density becomes very high before the
second phase transition of the fluxes, line A in figure
\ref{phase3dtop.ps}(b), occurs and consequently the fluctuations of the
instanton density measurements become very large in a larger region.

\section{Analytic and other approximations}
LAED contains both pure compact $U(1)$ and $\ZZ_2$ gauge theory. As both
of these theories have been studied thoroughly over the years, our aim is not
to make estimates for these models, but rather to treat their (numerical)
results as known and focus on the interaction of these two models in
LAED. To this end we give (analytical) approximations of some
characteristic quantities of the model. We subsequently discuss the
average action of unpierced plaquettes\footnote{The total average
action per plaquette is easily determined by this result and the flux
density.}, the flux condensation lines, the contours of constant flux
density in the region between the two flux condensation lines $A$ and
$B$ and the monopole/instanton density. We conclude this section with
a brief discussion of the approximations we made.
\label{approximations}

\subsection{The average action of unpierced plaquettes}
\label{aveaction}
To approximate the average action per unpierced plaquette,
$-\langle\cos{\tilde{F}}\rangle$, we split the parameter space of the
model into two regions, a region where the $\ZZ_2$ fluxes do not
condense and the region where they do.

In the region where the $\ZZ_2$ fluxes do not condense we approximate
the theory by a pure U(1) gauge theory (in the present context
considered to be given) and $-\langle\cos{\tilde{F}}\rangle$ is
approximated accordingly, i.e. we ignore the effect which the few
Alice fluxes have, that may be present. In the region where the fluxes
do condense and the flux density is large, we approximate the average
action of unpierced plaquettes by the average action of a single
plaquette.
The $U(1)$ link variables are irrelevant to plaquettes pierced by a
flux, as follows from formula (\ref{action2}). In the limit of a high
flux density the plaquettes which are not pierced by a flux become
isolated in the sense that the value of the $U(1)$ degrees of freedom
have almost no effect on the surrounding plaquettes. Thus we
can approximate $-\langle\cos{\tilde{F}}\rangle$ in the condensed phase
by:
\begin{equation}
\langle\cos{\tilde{F}}\rangle \approx
\frac{\int^{2\pi}_{0}\frac{d\tilde{F}}{2\pi} \cos{\tilde{F}}
e^{\frac{\cos{\tilde{F}}}{g^2}}}{\int^{2\pi}_{0}\frac{d\tilde{F}}
{2\pi}e^{\frac{\cos{\tilde{F}}}{g^2}}}=\frac{I_1(\frac{1}{g^2})}
{I_0(\frac{1}{g^2})}~,
\end{equation}
where the functions $I_0$ and $I_1$ are modified Bessel functions.

The difference between these two limits, the single plaquette and the
$U(1)$ limit, vanishes for large $g^2$. In four dimensions, for small
$g^2$, the fraction of pierced fluxes typically is very large in the
flux condensed phase. Thus we may expect that the two limits describe
the model for any value of $g^2$. In three dimensions there is no such
jump in the flux density and we expect an intermediate region, for
small $g^2$, to be present.
 
\begin{figure}[!htb]
\mbox{\psfig{figure=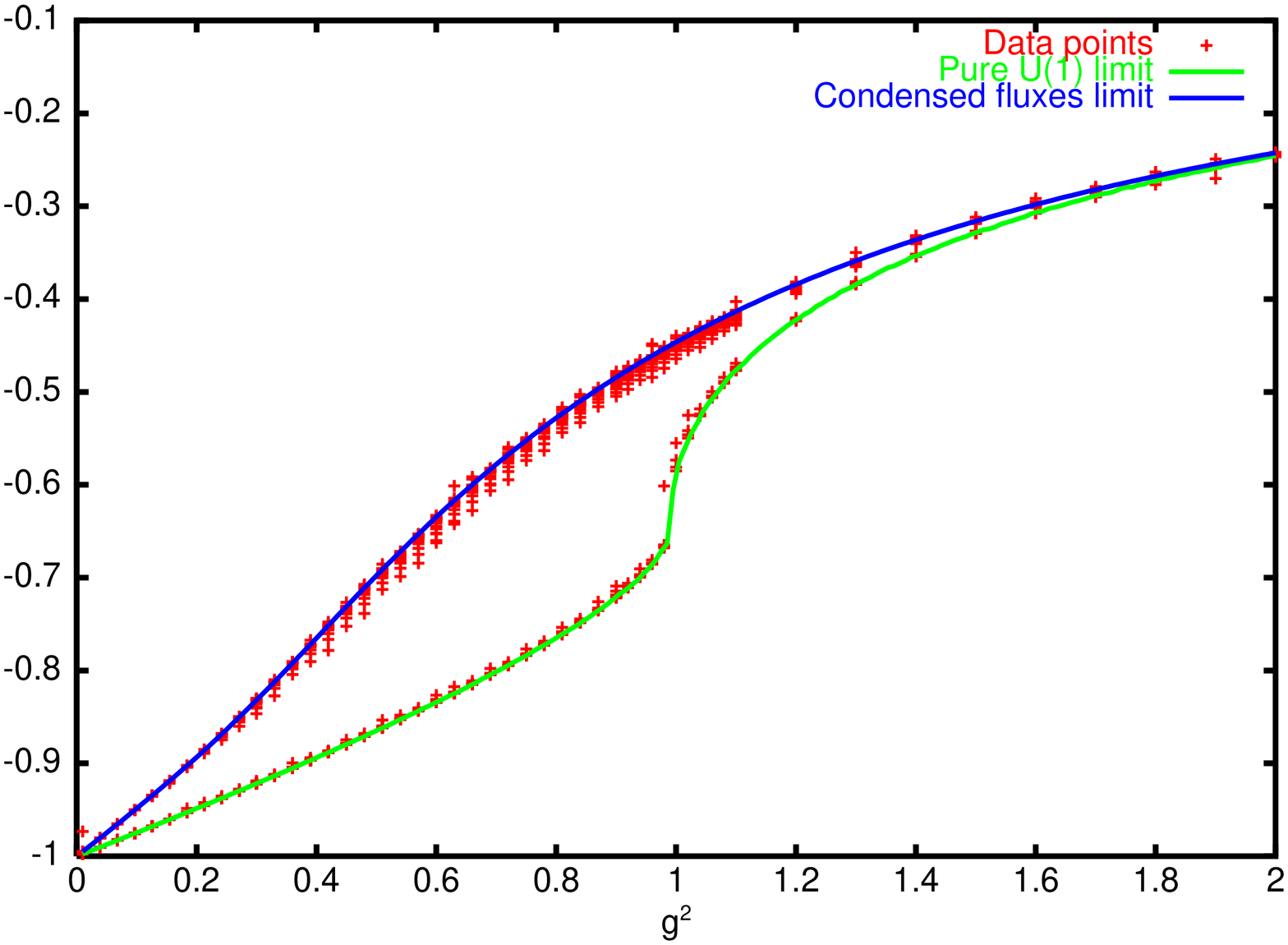,angle=270,width=7.9cm}}
\mbox{\psfig{figure=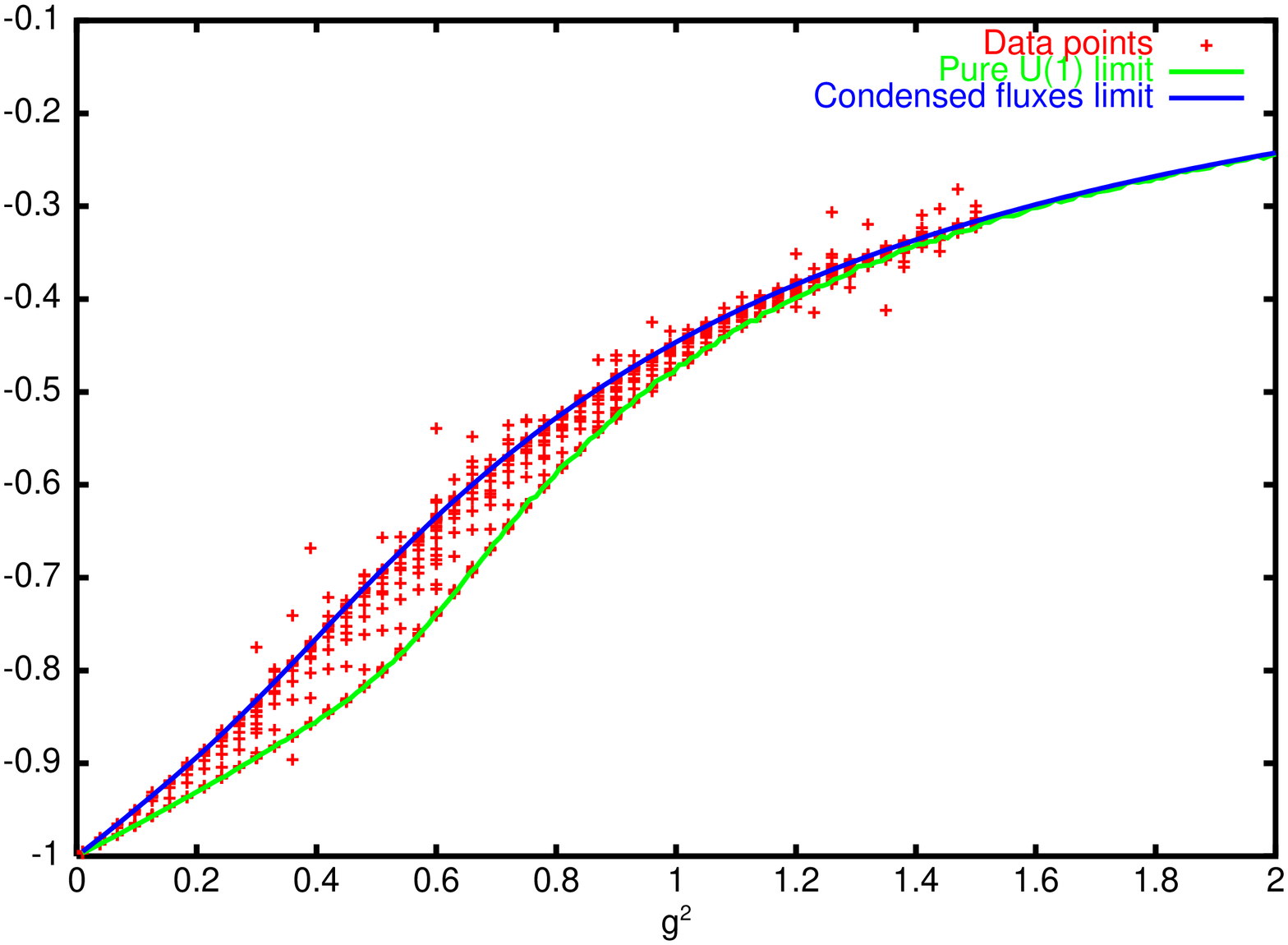,angle=270,width=7.9cm}}
\makebox[7.9cm][l]{\footnotesize{(a)}}
\makebox[7.9cm][l]{\footnotesize{(b)}}
\caption[somethingelse] 
{\footnotesize \\(a): The average plaquette action
$-\langle\cos{\tilde{F}}\rangle$ in four dimensions. All the data
points, i.e. including those corresponding to different values of
$m_f$ and $g^2$, lie either on the pure $U(1)$ line or (almost) on the
approximation for condensed fluxes phase. The division is so clear due
to the strong first order $\ZZ_2$ transition, i.e. in the flux
condensed phase the flux density is fairly high for small $g^2$.\\(b):
The average plaquette action $-\langle\cos{\tilde{F}}\rangle$ in three
dimensions. Here the transition from the one region to the other is
much more smooth, because the $\ZZ_2$ transition in three dimensions
is only of second order, also the pure $U(1)$ result deviates much
less from the flux condensed limit. The points outside the region
between the two limits are points where the flux density is very high,
implying that the fluctuations become very large.}
\label{cosF4d.ps}
\label{cosF3d.ps}
\end{figure}

In figure \ref{cosF4d.ps}(a) we plotted
$-\langle\cos{\tilde{F}}\rangle$ as a function of $g^2$ in four
dimensions. We see that the data splits up into two lines. Part of the
data points lie on the pure $U(1)$ line while the other part lies
(almost) on the single plaquette line. This strict separation of the
data points in these two sets is due to the strong first order
behaviour of the $\ZZ_2$ flux condensation for small $g^2$. We see that
each point is very well described by either the first or the second
approximation indeed.

In figure \ref{cosF3d.ps}(b) we plotted
$-\langle\cos{\tilde{F}}\rangle$ as a function of $g^2$ in three
dimensions. The two approximations now generate the boundaries between
which the data points lie. The fact that there is no clear division of
the data in two sets in three dimensions, is due to the fact that the
$\ZZ_2$ phase transition is of second order. The flux density grows
gradually across the transition region. That points appear also
outside the region bounded by the two approximations is due to very
large fluctuations when the flux density is high, i.e. when there are a
small number of unpierced plaquettes.

\subsection{The condensation lines of the Alice fluxes}
To approximate the location of the Alice flux condensation lines in
the parameter space of the model we make use of an action versus
entropy argument.  The weight factor of a configuration is determined
by $e^{S-I}$. The important quantity is the relative weight factor,
$e^{\Delta S-\Delta I}$, between configurations. Assuming that $S-I$
of the object that condenses, is additive with respect to the so
called background, we find that $\Delta S-\Delta I=S_{object}-
I_{object}$. Now typically the location of the critical point can be
approximated by $I_{object}= S_{object}$.

As we saw in figures \ref{phase4dtop.ps}(b) and \ref{phase3dtop.ps}(b)
there are two flux condensation lines in LAED. In the $\ZZ_2$ gauge
theory these are just each other mirror image. For finite $g^2$ this
symmetry between the two condensation lines is broken due to the
interactions with the $U(1)$ fields. We may still compare them, in the
sense that at the first transition line, B, the fluxes condense, while
at the other, A, the ``no-fluxes'' condense. The coupling between the
$\ZZ_2$ and the $U(1)$ fields manifests itself as follows: if a flux
is created then a piece of the $U(1)$ fields is ``eaten'' away, in the
sense that the $U(1)$ fields become irrelevant because they are
projected out and do not affect the action of the plaquettes
involved. This is an effect that we have to take into account, and as
we shall see, this can be done very accurately for the no-flux
condensation line, but only partially for the flux condensation line.
\\[2mm] {\bf The four dimensional case:}

First we determine the transition line of the ``no-flux'' condensate
with the help of the action versus entropy argument. When a no-loop
(i.e. a loop of no-flux) is created, the plaquettes through which it
pierces carry a $U(1)$ action. We determine the no-flux density and
will assume that the contributions of the $U(1)$ field of a plaquette
are independent of each other. We then approximate the location of the
condensation line by assuming that the average over the $U(1)$ degrees
of freedom in the relative weight factor for a plaquette is equal to
one. This gives us:
\begin{equation}
\ln{(\int^{2\pi}_{0}\frac{d\tilde{F}}{2\pi}e^{c_{nl}
+ \frac{m_f}{g^2}+ \frac{\cos{\tilde{F}}}{g^2}} )} = 0~,
\label{conden1}
\end{equation}
where $c_{nl}$ denotes the given value of the condensation point of
the no-loops in the pure $\ZZ_2$ gauge theory limit and we used
$\Delta I=-\frac{m_f}{g^2}-\frac{\cos{\tilde{F}}}{g^2}$ per
plaquette. We note that the value of $c_{nl}$ equals to minus the
value for the loops, $c_l$, as follows from the mirror symmetry of the
$\ZZ_2$ gauge theory, as we discussed at the end of section
\ref{implem}. From now on we will adopt the notation $c_{nl}=
-c_{4D}(\equiv-c_l)$.

\begin{figure}[!htb]
\mbox{\psfig{figure=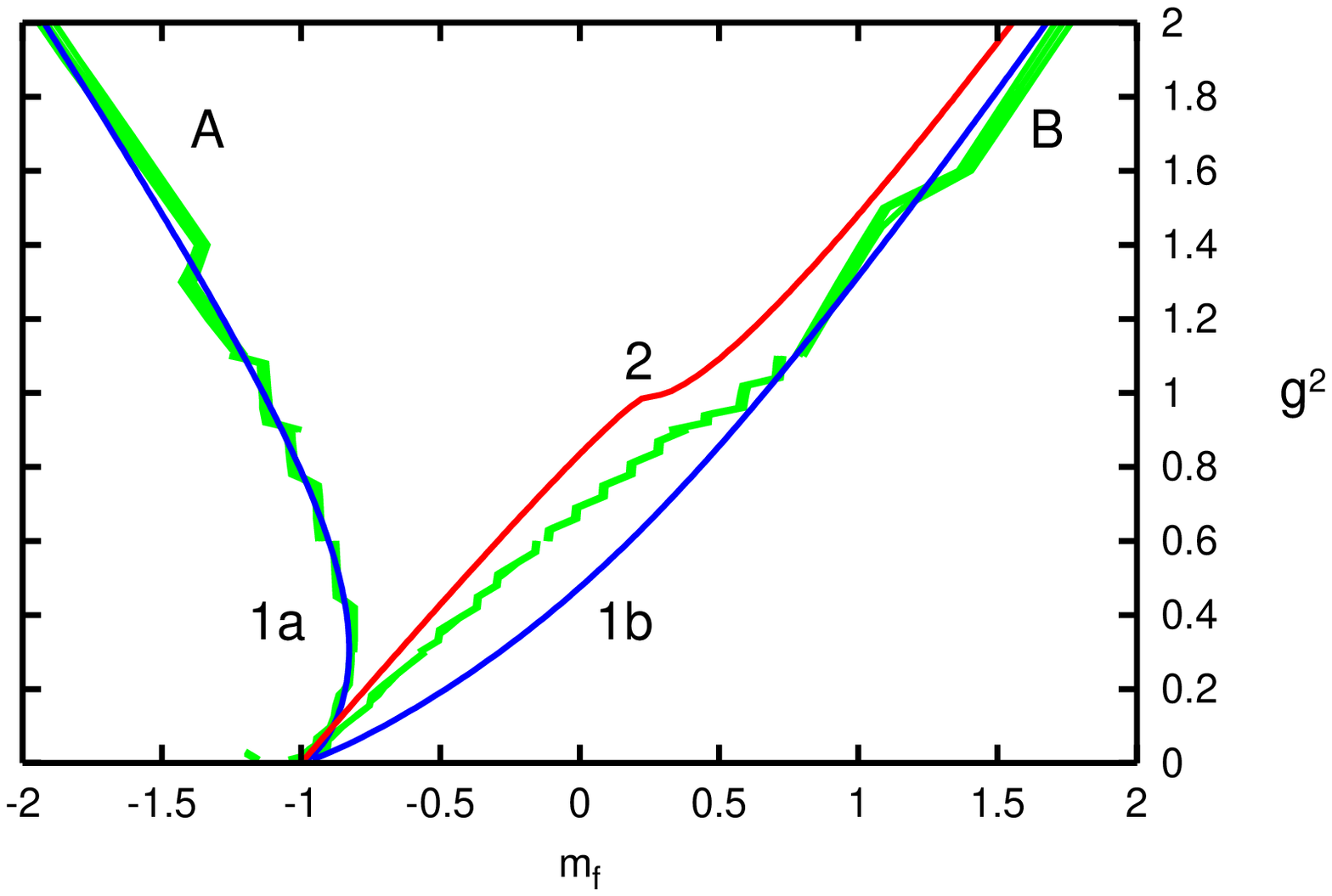,angle=270,width=7.9cm}}
\mbox{\psfig{figure=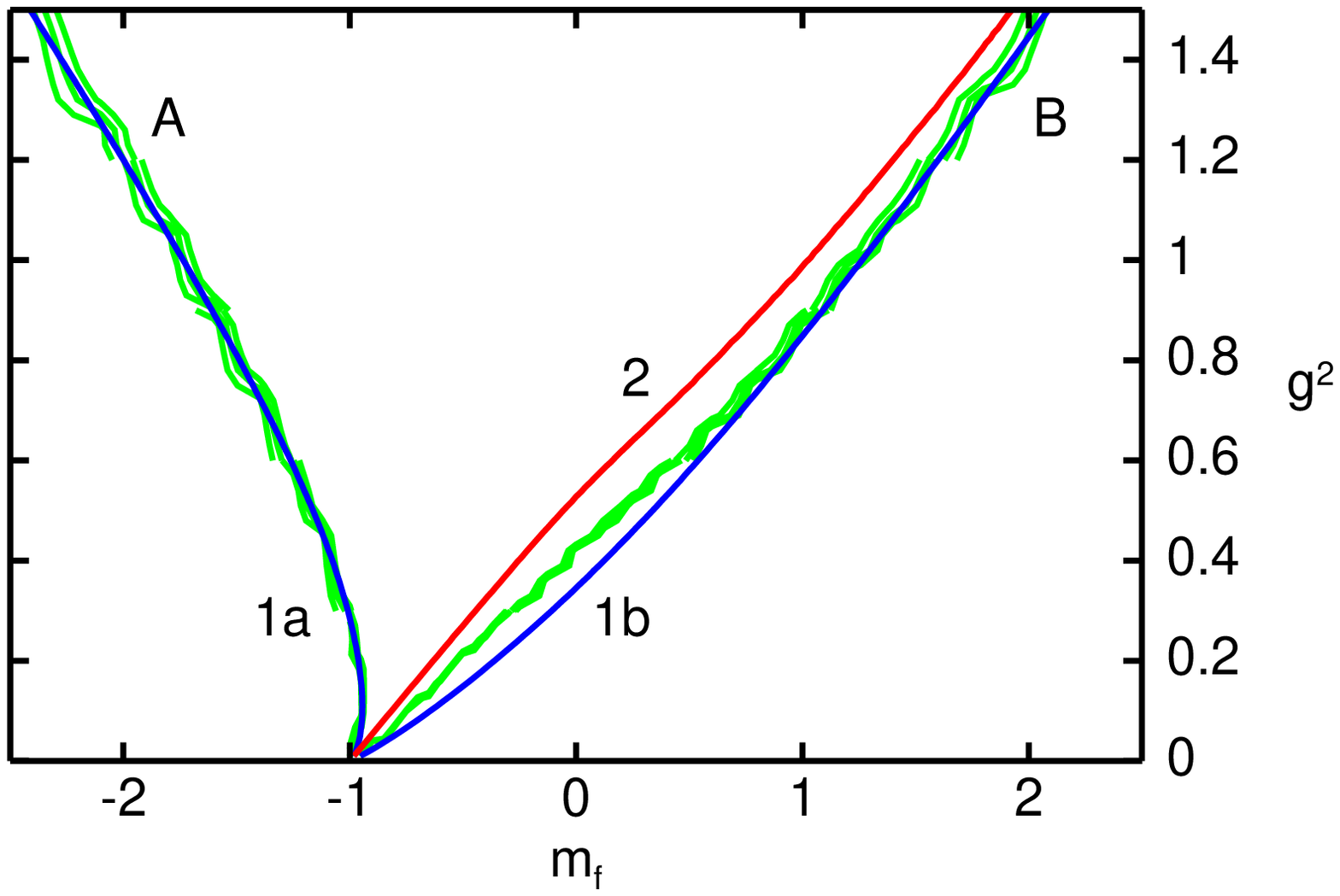,angle=270,width=7.9cm}}
\makebox[7.9cm][l]{\footnotesize{(a)}}
\makebox[7.9cm][l]{\footnotesize{(b)}}
\caption[somethingelse] 
{\footnotesize \\(a): A plot of the phase transition lines, A and B,
in four dimensions and of the approximations we made. The
approximation for the no-loop condensation line, 1a, is very good. For
$g^2>1$ the same approximation works also very good for the loop
condensation line, 1b, while the other approximation, 2, deviates in a
qualitatively expected way from the loop condensation line.\\(b): A
plot of the phase transition lines, A and B, in three dimensions and
of the approximations we made. The approximation for the no-flux
condensation line, 1a, is very good. For $g^2>0.6$ the same
approximation works also very good for the flux condensation line, 1b,
while the other approximation, 2, deviates in an expected way from the
flux condensation line.}
\label{fluxcond4d.ps}
\label{fluxcond3d.ps}
\end{figure}

Formula (\ref{conden1}) leads to the following equation for the
transition curve in the $(m_f,g^2)$ plane:
\begin{equation}
m_f = - g^2 c_{4D} - g^2 \ln{I_0(\frac{1}{g^2})}~.
\label{mf4d1}
\end{equation}
As can be seen in figure \ref{fluxcond4d.ps}(a) the approximation of
the no-loop condensation line is very good. 

We can try to do the same for the Alice loop condensation line. We use
again $ I_{object}= S_{object}$, but are now not able to include all
contributions. The entropy and action contribution of the loop are
clear, one thing that changes in equation (\ref{mf4d1}) is the sign in
front of the first term on the r.h.s.. The problem is a reliable
estimate of the $U(1)$ contribution. Obviously we may no longer assume
that the $U(1)$ contribution of each plaquette is independent. On the
other hand it is known that the correlation length decreases
exponentially in the confining phase, which implies that we should
expect this approximation to still work if $g^2>g^2_c\approx 1$.

We can also approximate the Alice loop condensation line in a slightly
different way, where we use the contribution to the action of the
$U(1)$ fields as given by the pure $U(1)$ theory and ignore the change
in the entropy due to the $U(1)$ fields. For the action we then take:
\begin{equation}
 I_{object} = (\frac{m_f}{g^2} +
 \frac{\langle\cos{\tilde{F}}\rangle}{g^2})~,
\end{equation}
with $\langle\cos{\tilde{F}}\rangle$ the average of $\cos{\tilde{F}}$
for given $g^2$ and is equal to $\langle\cos{F}\rangle$ of pure $U(1)$
gauge theory as follows from the previous section (which is evaluated
numerically and in the present context considered as given). This
leads to the following approximation for the position of the
condensation line for the loops:
\begin{equation}
m_f = g^2 c_{4D} - \langle\cos{\tilde{F}}\rangle~.
\label{mf4d2}
\end{equation}
We note that in the pure $\ZZ_2$ limit, the second term on the
r.h.s. of equations (\ref{mf4d1}) and (\ref{mf4d2}) becomes zero and
that $c_{4D}$ and its three dimensional analogue $c_{3D}$ follow from
pure $\ZZ_2$ gauge theory results as mentioned before. In fact, they
are even known analytically \cite{Balian}. In the limit of $g^2\to 0$ the
only state that is allowed, is the global minimum, which means that the
condensation lines need to go to $m_f=-1$ for $g^2\to 0$. This is true
for both approximations.

In figure \ref{fluxcond4d.ps}(a) we have plotted the approximations
for the condensation lines in four dimensions and some specific height
lines, which characterise the position of the phase transitions.  We
see that the approximation of the condensation of the no-loops is very
good. For $g^2>1$ the same method works also very well for the loop
condensation line. The other approximation for the loop condensation
line does not work as well, but we qualitatively understand why.
\\[2mm]{\bf The three dimensional case:}

In three dimensions we follow the same strategy. We repeat
the arguments given for the four dimensional case, leading to exactly
the same equations (\ref{mf4d1}) and (\ref{mf4d2}), where we only have
to replace the four dimensional quantities by their three dimensional
counterparts. In particular $c_{4D}$ is replaced by $c_{3D}$ and
$\langle\cos{\tilde{F}}\rangle_{4D}$ is replaced by
$\langle\cos{\tilde{F}}\rangle_{3D}$.

In figure \ref{fluxcond3d.ps}(b) we plotted the resulting condensation
lines for the three dimensional theory. The plot shows some specific
height lines which characterise the phase transitions as well as the
approximations for the lines where the phase transitions should
occur. Again we find that the approximation for the no-flux
condensation line is very good. The approximation of the analogue of
equation (\ref{mf4d1}) is very good for larger values of $g^2$, whereas
the deviation of the other approximation to the flux condensation line
is qualitatively understood.

\subsection{Contours of constant flux density}
 \label{heightlines} In this subsection we will approximate the flux
 density in the region between the two flux condensation lines, by
 assuming that in this region the correlation lengths of both fields are
 zero, so that it suffices to look at the single plaquette.

This means that we get the same answer for the three and four
dimensional case. The fraction of plaquettes being pierced by an Alice
flux, $\rho_f$, can be approximated by:
\begin{equation}
\rho_f\approx \frac{e^{ s_f - I_f}}{e^{ s_f - I_f} + e^{ s_{nf}}\int^{2\pi}_{0} \frac{d\tilde{F}}{2\pi}~e^{- I_{\tilde{F}}}}~.
\end{equation}
Using $ s_f= s_{nf}$ and $I_f=\frac{m_f}{g^2}$ this gives:
\begin{equation}
m_f = g^2\ln{\frac{1-\rho_f}{\rho_f}} - g^2
\ln{\int^{2\pi}_{0}\frac{d\tilde{F}}{2\pi}~e^{- I_{\tilde{F}}}}~,
\end{equation}
which leads to:
\begin{equation}
m_f = g^2\ln{\frac{1-\rho_f}{\rho_f}}- g^2 \ln{I_0(\frac{1}{g^2})}~.
\label{height2}
\end{equation}
Note that in the limit $g^2\to 0$ we find that all the height lines
meet at $m_f=-1$, just as one should expect, whereas in the $\ZZ_2$
limit one obtains that $\frac{m_f}{g^2} =
\ln{\frac{1-\rho_f}{\rho_f}}$.

\begin{figure}[!htb]
\mbox{\psfig{figure=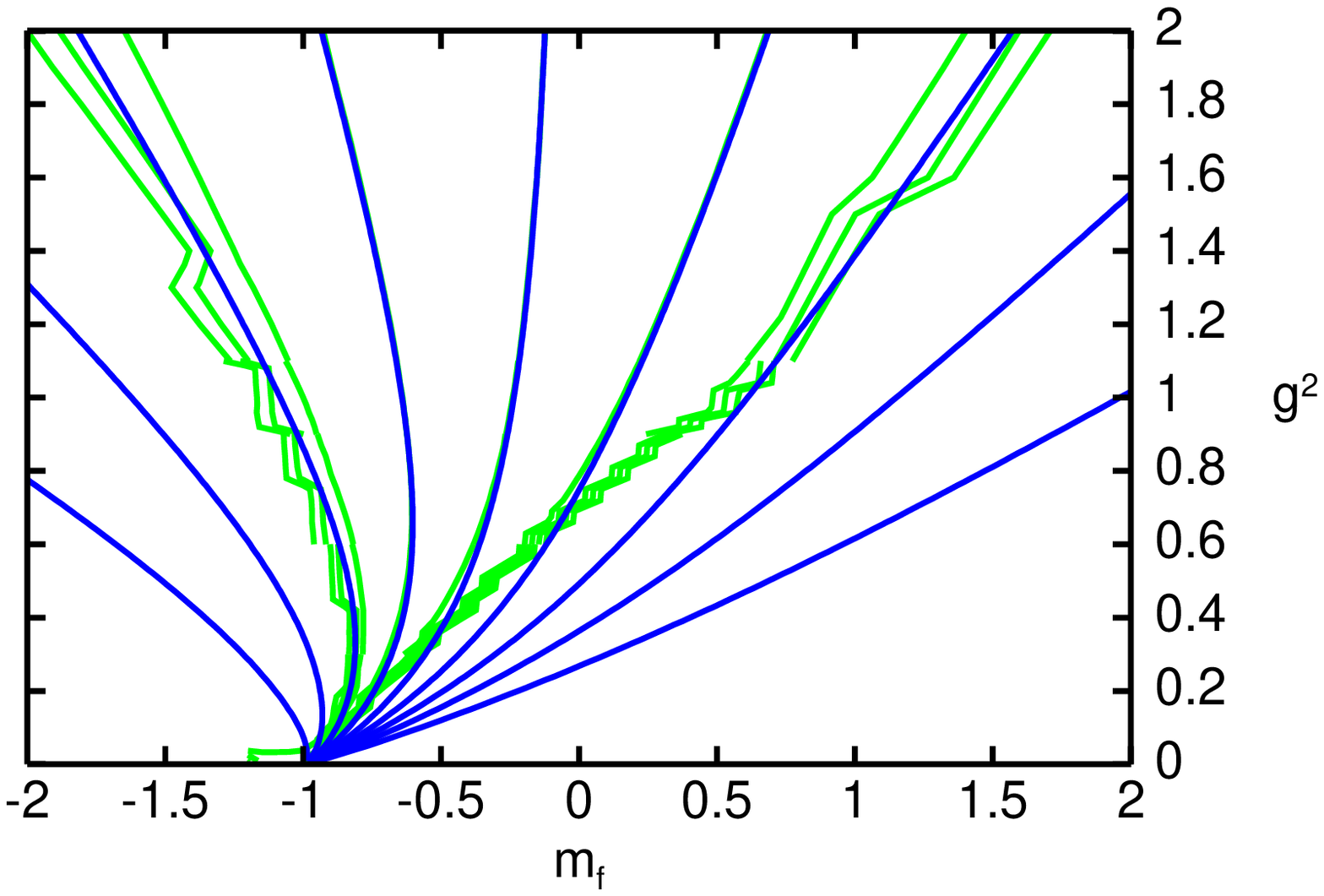,angle=270,width=7.9cm}}
\mbox{\psfig{figure=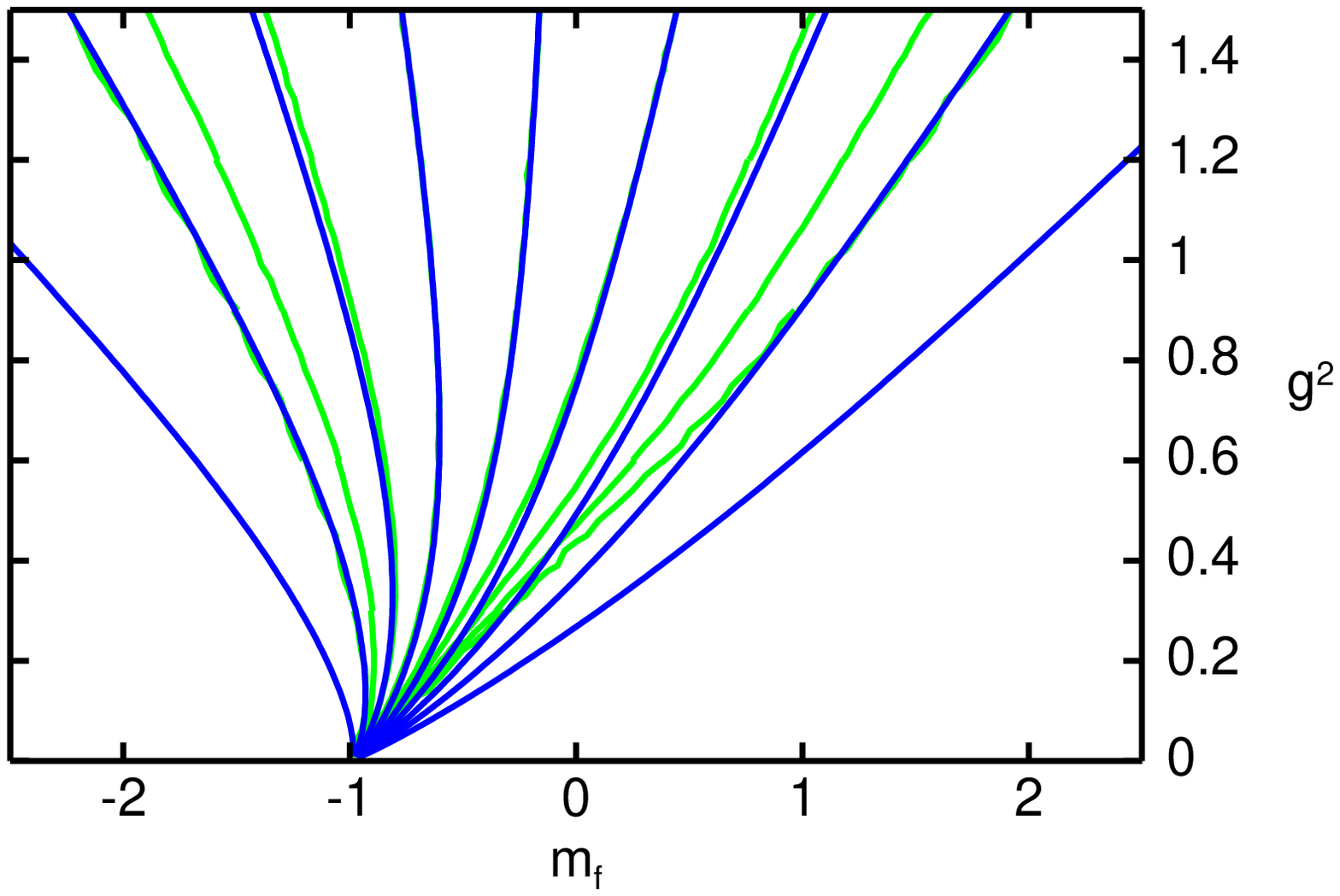,angle=270,width=7.9cm}}
\makebox[7.9cm][l]{\footnotesize{(a)}}
\makebox[7.9cm][l]{\footnotesize{(b)}}
\caption[somethingelse] 
{\footnotesize \\(a): Contour lines of the flux density in four
dimensions and their approximations. We plotted from, left to right,
the height lines: $0.9,0.8,\cdots ,0.2,0.1$. The approximations for
the height lines $0.6,\cdots,0.4$ are perfect up to the point where
they reach the condensation line.\\(b): Contour lines of the flux
density in three dimensions and their approximations. We plotted from,
left to right, the height lines: $0.9,0.8,\cdots ,0.2,0.1$. The
approximations for the height lines $0.7,\cdots,0.3$ are very good up
to the point where they reach the condensation line.}
\label{hightlines4d.ps}
\label{hightlines3d.ps}
\end{figure}

In four dimensions, within the region of the two condensation lines,
which is the region we are probing, our approximation works very well,
see figure \ref{hightlines4d.ps}(a). In three dimensions the
approximation does not work in the whole region, but works very well
between the height lines $0.7$ and $0.3$, see figure
\ref{hightlines3d.ps}(b).

The approximation of the flux density, equation (\ref{height2}), can be
split into two parts.  The first term on the right hand side is due to
the $\ZZ_2$ degrees of freedom. In the $\ZZ_2$ limit this term can be
compared with pure $\ZZ_2$ gauge theory, which we did not use as input
in this estimate. The second term on the right hand side is due to the
$U(1)$ degrees of freedom. Moving away from the $0.5$ height line
makes the approximation of $\ZZ_2$ term less good while moving from
the $0.9$ height line to the $0.1$ height line makes the $U(1)$ term
less good. The validity of the $U(1)$ term can be seen by fitting the
$\ZZ_2$ part of the approximation with results from pure $\ZZ_2$ gauge
theory. This gives a perfect fit for all values $g^2$ for a high flux
density, but as one expects, fails in the region of low flux density
and small $g^2$.

\subsection{The monopole/instanton density}
\label{monopole}
In this subsection we will approximate the monopole/instanton
density. In the phase where the Alice fluxes do not condense the
monopole condensation line and height lines are easily understood. In
this phase there are almost no fluxes, and ignoring these the model
becomes a pure $U(1)$ theory and on expects the monopole density to
behave accordingly, allowing us to use the known numerical results.
\begin{figure}[!htb]
\mbox{\psfig{figure=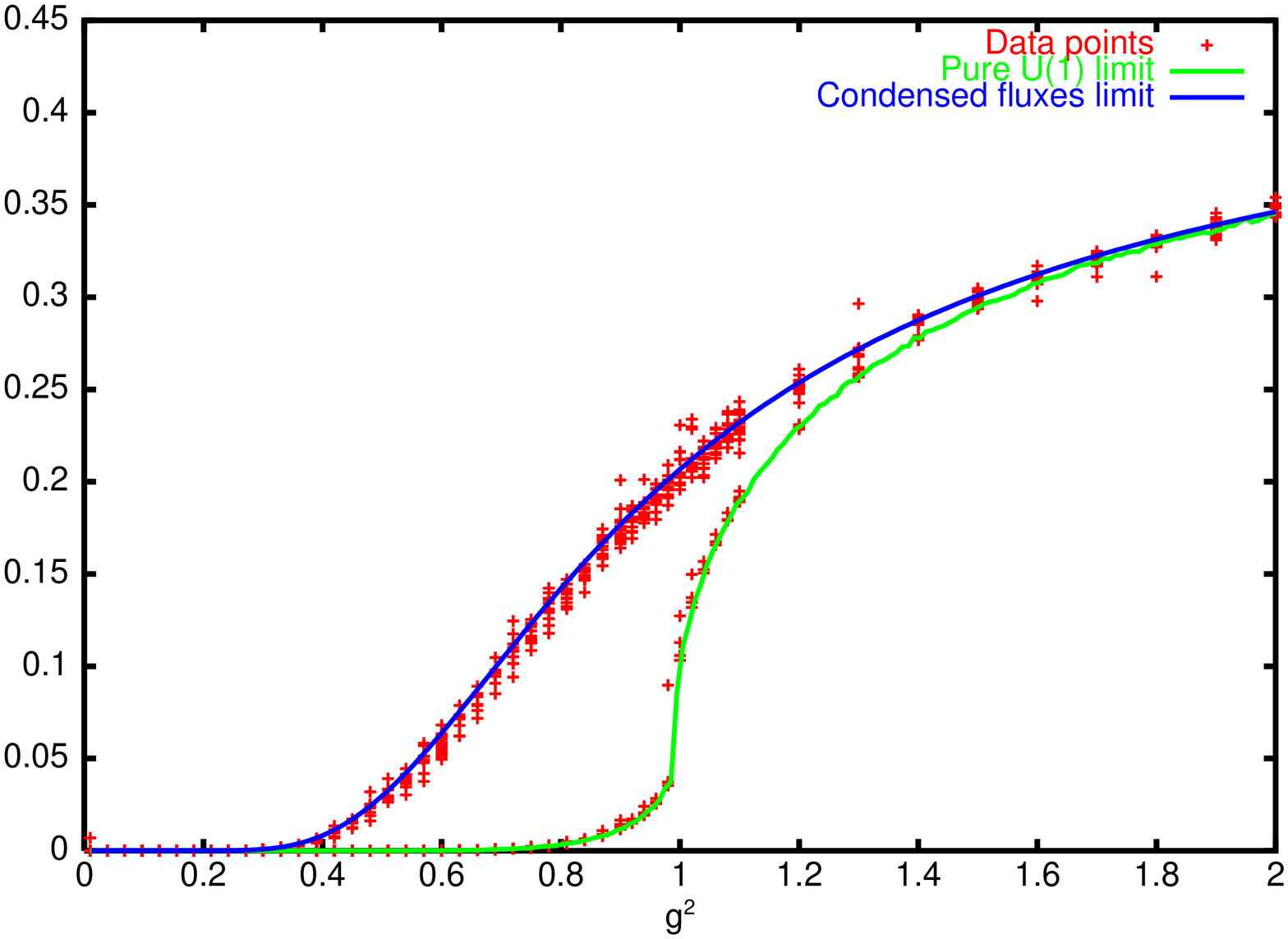,angle=270,width=7.9cm}}
\mbox{\psfig{figure=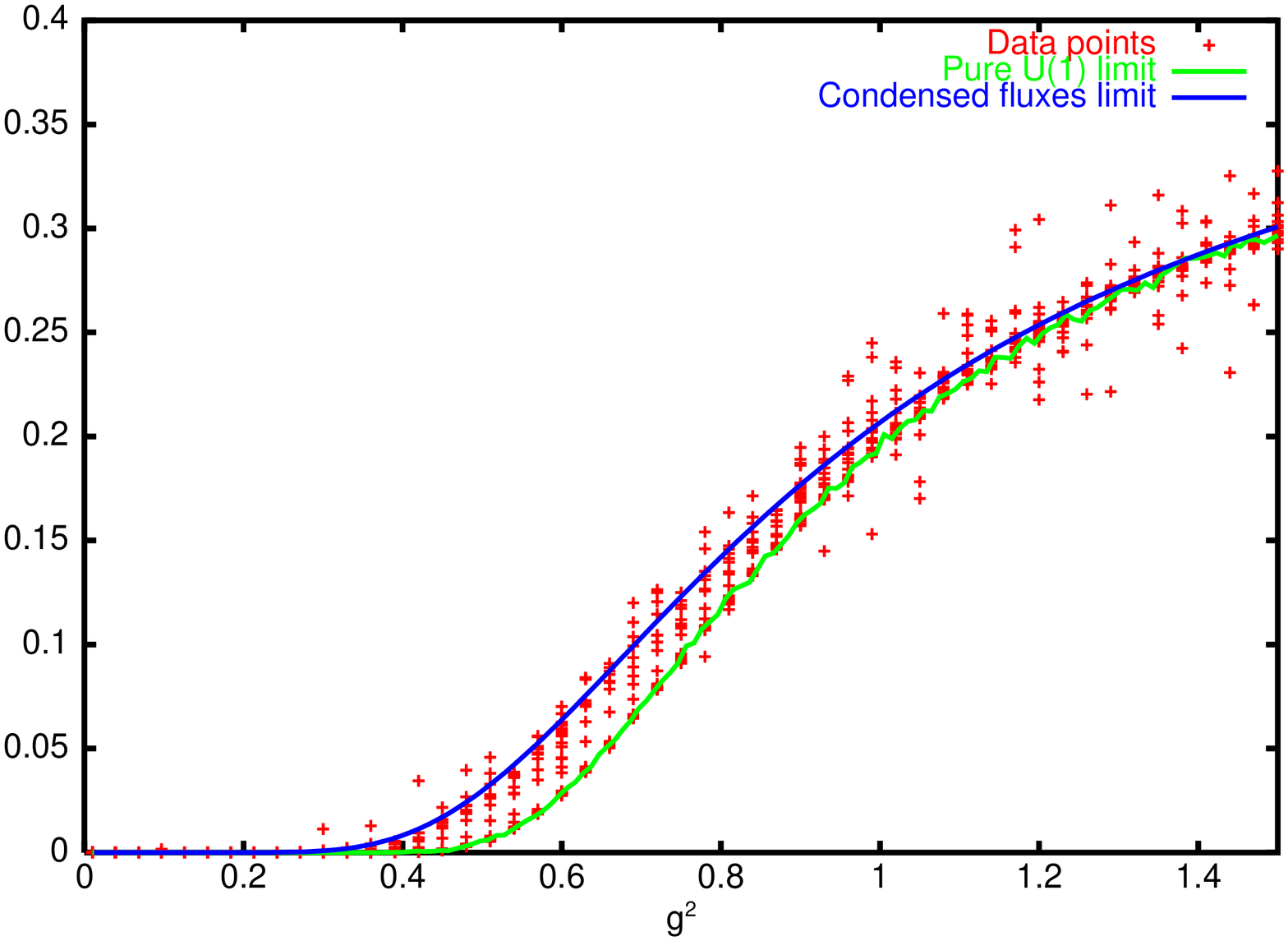,angle=270,width=7.9cm}}
\makebox[7.9cm][l]{\footnotesize{(a)}}
\makebox[7.9cm][l]{\footnotesize{(b)}}
\caption[somethingelse] 
{\footnotesize \\(a): A plot of the monopole density. Just as in
figure \ref{cosF4d.ps}(a) all the data points, i.e. including those
for different values of $m_f$ and $g^2$, perfectly match the two
different approximations. The monopole density is or equal to the pure
$U(1)$ monopole density or is (almost) equal to the approximation for
the condensed fluxes phase.\\(b): A plot of the instanton density. The
instanton density lies between the two different approximations. That
the data does not jump from one line to the other line is due to the
fact that the $\ZZ_2$ transition is much softer in three
dimensions. The points outside the region between the two limits are
points where the flux density is very high, implying that the
statistics is bad.}
\label{monopole4d.ps}
\label{monopole3d.ps}
\end{figure}

In the phase where the Alice fluxes do condense we may approximate the
monopole density by the monopole density of a single cube.  That this
can be done follows basically from the results of sections
\ref{aveaction} and \ref{heightlines}. The cubes not pierced by any
$\ZZ_2$ flux are in the condensed phase isolated in the sense that the
$U(1)$ degrees of freedom of the links have hardly any effect on the
surrounding plaquettes. This makes it safe to use the single cube
approximation in the phase where the fluxes have condensed.

We determined the single cube density by using random link values,
with which we determined the energy of the cube, the charge inside the
cube and the entropy of the configuration. With this information we
calculated the monopole density for different values of $g^2$ and
compared it with the data points we found. This approximation is the
same for the three and four dimensional model, though in three
dimensions these are of course instantons.

Just as in section \ref{aveaction} one expects the two approximations
to describe the model very well in four dimensions, but in three
dimensions one expects an intermediate region. This is exactly what we
find, see figures \ref{monopole4d.ps}(a) and
\ref{monopole3d.ps}(b). Again we note that the points outside the
region bounded by the two approximations are points where the flux
density is very high, i.e. the fluctuations become very large.

\subsection{Discussion}
\label{disofres}
The approximations we made in the last few sections describe the model
fairly well. In four dimensions the approximations work extremely
well. The phase with condensed fluxes can apparently be
understood as a phase where the correlation lengths of the fields are
vanishingly small. In three dimensions the division of the phase space
is not as clear, but our approximation of the
height lines of the flux density does imply a region where the
correlation length of both fields is also vanishingly small.
If the Alice fluxes do not condense the theory is very well described
by a pure compact $U(1)$ gauge theory. 

\begin{figure}[!htb]
\mbox{\psfig{figure=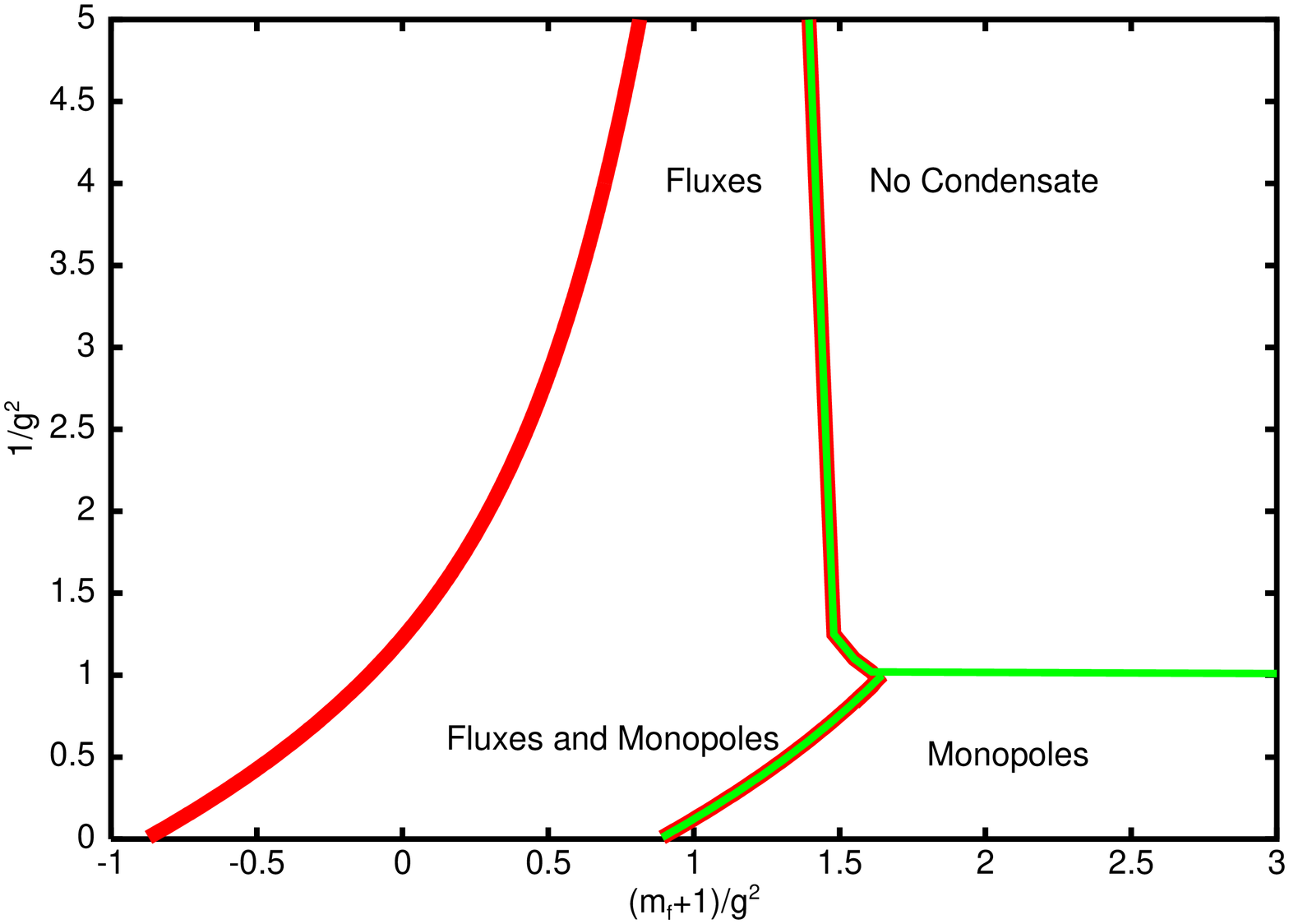,angle=270,width=7.9cm}}
\mbox{\psfig{figure=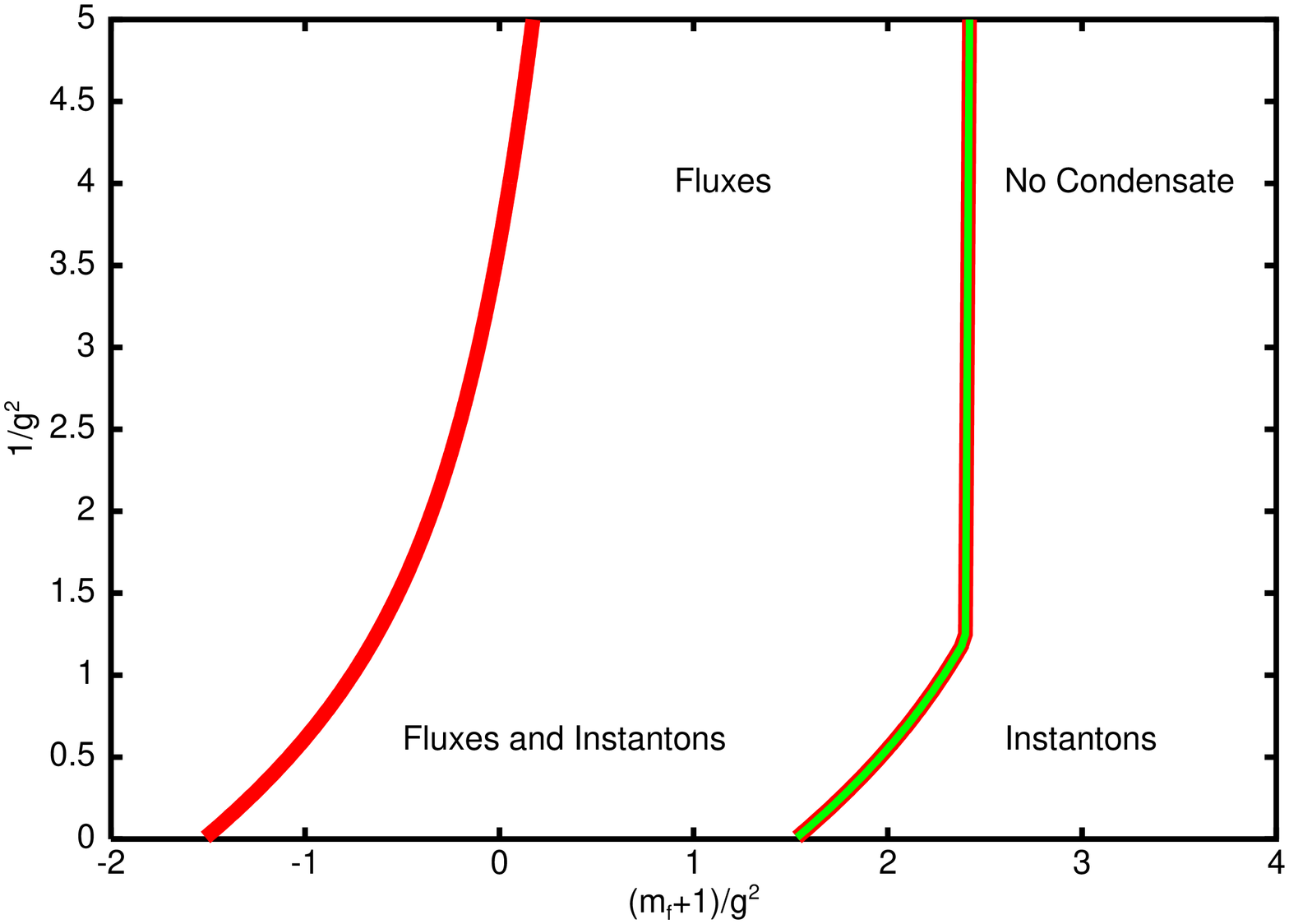,angle=270,width=7.9cm}}
\makebox[7.9cm][l]{\footnotesize{(a)}}
\makebox[7.9cm][l]{\footnotesize{(b)}}
\caption[somethingelse] 
{\footnotesize The phase diagrams of four, (a), and three, (b),
dimensional LAED in the new parameters. The details and implications
are explained in the text.}
\label{phasediagram4Dhand.ps}
\end{figure}

As mentioned before, the fact that all the contour lines of the flux
density come together at $m_f=-1$ for $g^2\to0$, does not mean that
the phase transition becomes or stays first order. It is mainly due to
the choice of parameters that all the contour lines of the flux
density come together. If one uses the in some sense more natural
parameters $(m_f+1)/g^2$ and $1/g^2$, it is not at all clear that this
will happen.  This is illustrated in in figure
\ref{phasediagram4Dhand.ps}, where we have plotted the phase diagram
of the model in terms of the conventional parameters.  The
crossover transitions are not marked, they are associated to regions
with different condensates not separated by a phase transition
line. Although there is a second flux transition
line, the ``no-flux'' condensation, there is no monopole/instanton
transition at this point. We deduce this from the results of section
\ref{monopole}. The fact that we are not able to determine the
monopole/instanton density there is due to the fact the
fluctuations are very large in that region of parameter space. However
one would expect the single cube approximation of section
\ref{monopole} also to be valid in that region of parameter space.

The position of the monopole transition line, see figure
\ref{phasediagram4Dhand.ps}(a) is also following from the results of
section \ref{monopole}. We pointed out that the monopole data splits
up into two regions, the regions where the fluxes have or have not
condensed. This means that the $U(1)$ monopole transition line splits
up and follows the (first) flux transition line. We have drawn it all
the way along this flux condensation line, but it is not yet clear
whether there is always a monopole transition. For $g^2\to 0$ and
$g^2\to\infty$ the difference in the monopole density between the two
regions becomes smaller and smaller.

To some extend the same is true for the instanton density, see figure
\ref{phasediagram4Dhand.ps}(b). Although in that case there is an
intermediate region, see section \ref{monopole}. In this region the
instanton density grows with increasing flux density, and since in
this region the flux density has a transition one would expect also
the instanton density to show a transition. The data also appears to
imply this, but is not shown here. Again it is not clear what happens
to this transition in the limits of $g^2\to 0$ and $g^2\to\infty$. In
these limits the difference of the instanton density between the
regions where the fluxes have or have not condensed goes to zero.

\section{Conclusions and outlook}
We have studied Alice electrodynamics on a lattice, with a model that
allows the formation of magnetic monopoles and Alice fluxes. It
includes the usual Wilson lattice action for the $U(1)$ gauge theory
and has an extra bare mass term for Alice fluxes. This term suffices to
reach all four phases of Alice electrodynamics given in table
\ref{phases}.

We have determined the regions in phase space corresponding to the
four different phases of LAED and presented results on some measurable
quantities; the monopole density, the flux density and
$\langle\cos{\tilde{F}}\rangle$. We then approximated the locations of
the flux and the so called no-flux condensation line in the phase
diagram of the model, both in three and four dimensions. These
approximations worked very well except for the flux condensation line
for small values of the gauge coupling. The other approximations we
made also all work quite well, with the remark that in three
dimensions there is an intermediate region which we have not yet
investigated. We successfully compared our numerical results with
approximations of the flux density between the flux and the no-flux
condensation line, the monopole/instanton density,
$\langle\cos{\tilde{F}}\rangle$ and the position of the monopole
condensation line. The monopole condensation becomes a crossover in
the region where the Alice fluxes are condensed. In section
\ref{disofres} we gave the resulting phase diagrams.

It would be interesting to examine the fate of the phase transitions
in the monopole and instanton density induced by (first) condensing
Alice fluxes for small and large values of $g^2$. For small values of
$g^2$ it is also not clear if the two flux transitions merge or not in
the parameter space with the coordinates $(m_f+1)/g^2$ and $1/g^2$.
In a forthcoming paper we will address interesting questions
concerning the screening versus confinement of charges and/or magnetic
monopoles in the various phases of the theory.

We thank Jan Smit and Jeroen Vink for valuable advice and support
related to lattice gauge theories. This work was partially supported
by the ESF COSLAB program.

\end{document}